\documentclass{jfm_cup}
\usepackage{graphicx}
\usepackage{epstopdf, epsfig}
\usepackage{hyperref}
\usepackage[usenames,dvipsnames,svgnames,table]{xcolor}
\usepackage{libertine}
\usepackage[utf8]{inputenc}
\usepackage{newtxmath}
\usepackage{pgfplots}
\usepackage{natbib}
\usepackage{microtype}
\usepackage{bm}
\usepackage[normalem]{ulem} 

\hypersetup{
colorlinks=true,
linkcolor=blue,
urlcolor=blue,
filecolor=blue,
citecolor=blue
}

\makeatletter

\patchcmd{\NAT@citex}
  {\@citea\NAT@hyper@{%
     \NAT@nmfmt{\NAT@nm}%
     \hyper@natlinkbreak{\NAT@aysep\NAT@spacechar}{\@citeb\@extra@b@citeb}%
     \NAT@date}}
  {\@citea\NAT@nmfmt{\NAT@nm}%
   \NAT@aysep\NAT@spacechar\NAT@hyper@{\NAT@date}}{}{}

\patchcmd{\NAT@citex}
  {\@citea\NAT@hyper@{%
     \NAT@nmfmt{\NAT@nm}%
     \hyper@natlinkbreak{\NAT@spacechar\NAT@@open\if*#1*\else#1\NAT@spacechar\fi}%
       {\@citeb\@extra@b@citeb}%
     \NAT@date}}
  {\@citea\NAT@nmfmt{\NAT@nm}%
   \NAT@spacechar\NAT@@open\if*#1*\else#1\NAT@spacechar\fi\NAT@hyper@{\NAT@date}}
  {}{}

\makeatother

\makeatletter
\newcommand{\average}[2][0]{{%
  \mspace{#1mu}%
  \overline{\mspace{-#1mu}\average@check#2\relax}%
}}
\newcommand\average@check[1]{%
  #1\@ifnextchar_{\average@sub}{}%
}
\newcommand{\average@sub}[2]{% #1 is _
  _{#2}\mspace{-2mu}\aftergroup\average@compensate
}
\newcommand{\average@compensate}{\mspace{2mu}}
\makeatother
\defcitealias{companion}{GMS}

\newcommand{\vect}[1]{\boldsymbol{#1}}
\newcommand{\matrgr}[1]{\vect{#1}}
\newcommand{\matr}[1]{\mathsfbi{#1}}

\newcommand\Fr{\mbox{\textit{Fr}}} % Froude number
\newcommand\im{\mathrm{i}\mkern1mu} % imaginary unit

\DeclareMathOperator{\spn}{span}

\title{Ill posedness in shallow multi-phase debris flow models}
\shorttitle{Ill posedness in multi-phase debris flow models}

\shortauthor{J. Langham, X. Meng, J. P. Webb, C. G. Johnson, J. M. N. T. Gray}
\author{Jake Langham\aff{1}
  \corresp{\email{jacob.langham@manchester.ac.uk}},
  Xiannan Meng\aff{2}
  \corresp{\email{xiannan.meng@dlmu.edu.cn}},
  Jamie P. Webb\aff{1},
  Chris G. Johnson\aff{1},
  J. M. N. T. Gray\aff{1}}

\affiliation{\aff{1}Department of Mathematics and Manchester Centre for Nonlinear Dynamics, University of Manchester, Oxford Road, Manchester M13 9PL, UK
\aff{2}Transportation Engineering College, Dalian Maritime University, Dalian 116026, PR China
}

\begin{document}

\maketitle

\begin{abstract}
Depth-averaged systems of equations describing the motion of fluid--sediment
mixtures have been widely adopted by scientists in pursuit of models that
can predict the paths of dangerous overland flows of debris.
As models have become increasingly sophisticated,
many have been developed from a multi-phase perspective in which
separate, but mutually coupled sets of equations govern
the evolution of different components of the mixture.
However, this creates the opportunity for the existence of pathological
instabilities stemming from resonant interactions between the phases.
With reference to the most popular approaches, analyses
of two- and three-phase models are performed, which demonstrate that they are more often than not
ill posed as initial value problems over physically relevant parameter
regimes -- an issue which renders them unsuitable for scientific
applications.
Additionally, a general framework for detecting ill posedness in 
models with
any number of phases is developed.
This is used to show that small diffusive terms in the equations for momentum
transport, which are sometimes neglected, can reliably eliminate this issue. 
Conditions are derived for the regularisation of models in this way, but they
are typically not met by multi-phase models that feature diffusive terms.
\end{abstract}

\section{Introduction}
\label{sec:intro}
Debris flows are large-scale gravity currents that are formed on hillslopes when
water entrains and mixes with rocks, mud, and other natural detritus. Despite
their daunting physical complexity, the threat they pose to human life
\citep{Dowling2014} motivates ongoing efforts to develop detailed model
descriptions of them, for the purposes of hazard prediction and risk
assessment~\citep{Hutter1994,Iverson1997,Trujillo2022}. 

The commonest class of available models are variations on the classical
depth-averaged shallow water equations, re-derived to incorporate physical
effects particular to debris flows, such as non-Newtonian stresses, buoyancy and
pore water pressure.  Early approaches considered flows to be sufficiently
homogeneous that the mass and momentum of fluid and submerged debris could be
lumped together into a single continuous phase, subject to bulk conservation
laws
\citep{Savage1989,Macedonio1992,Iverson1997,Fraccarollo2000,Iverson2001,Christen2010}.
While this perspective is sometimes justified, it cannot fully account for
important phenomena that arise from interactions between different components
within the flow, such as changes in the debris composition due to dilation and
particle size segregation, which can have a profound effect on the
dynamics~\citep{Hutter1994,Iverson1997,Berti2000,Mccoy2010,Johnson2012}.
Consequently, some models have included an equation for the transport of an
additional phase of solid particles within the flow, enabling solutions to
develop compositional variations that may in turn, affect the local fluid
rheology~\citep{Takahashi1992,Shieh1996,Brufau2000}. This approach may be
augmented by introducing coupled equations for the evolution of the vertical
distribution of solids~\citep{Kowalski2013}, or the basal pore-fluid
pressure~\citep{Iverson2014,George2014}.
A related strategy is to consider the transport of two or more species of
granular material, while neglecting the presence of a carrier
fluid~\citep{Gray2010}.  When combined with velocity shear through an assumed
vertically segregated flow column and frictional dependence on particle size,
this can likewise capture complex phenomena that are inaccessible to the
simplest models, including thickened fronts that dam the
flow~\citep{Denissen2019} and spontaneous finger
formation~\citep{Woodhouse2012,Baker2016}.

Truly `multi-phase' systems take a step further by disaggregating the momentum
dynamics of the different phases, thereby permitting the forces acting on each
constituent to be modelled
separately~\citep{Pitman2005,Pelanti2008,Pailha2009,Pudasaini2012,Bouchut2016,Li2018,Pudasaini2019,Meyrat2022,Meng2022,Meng2024}.
Model development in this final category is ongoing and promises to deliver the
most faithful realisation of debris flow physics within the depth-averaged
framework, particularly when there is significant separation of phases within
the flow.

However, the specification of separate momentum equations for multiple flow
phases can introduce a fundamental pathology into depth-averaged models, causing
them to no longer reflect the behaviour of the underlying physical system.  For
example, when a second fluid layer is added to the classical shallow water
equations, they cease to be unconditionally strictly
hyperbolic~\citep{Ovysannikov1979}, leaving the system ill-posed as an initial
value problem when the flow is in certain conditions. The underlying reason for
this is that buoyancy-mediated coupling between the two layers introduces a
linear instability with a growth rate that diverges to infinity in the limit of
high wavenumber perturbations. 
A practical consequence of this is that time-dependent simulations of the
system in these conditions are guaranteed to be mesh dependent. Therefore,
much attention has been given towards developing physically defensible
methods which locally amend this model or otherwise drive solutions away from
non-hyperbolic regimes~\citep[see e.g.][]{Castro2001,Sarno2017,Krvavica2018,CastroDiaz2023}.

\begin{figure}
    \centering
    \includegraphics[width=0.8\textwidth]{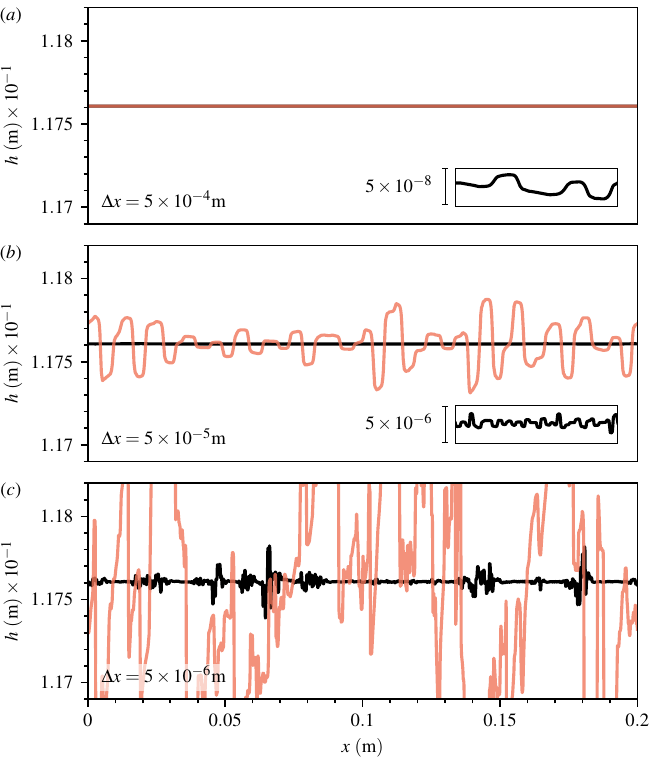}%
    \caption{Demonstration of ill posedness, using the model of \cite{Meng2022}.
    Snapshots of total flow depth ($h = H_f+H_s$ in the notation of
    \S\ref{sec:mengmodel}) at times $t=1\mathrm{s}$ (black) and
    $2\mathrm{s}$ (red) are plotted
    for numerical simulations of an initially uniform steady flow in a periodic
    domain of length $0.2\mathrm{m}$, subject to a small noisy
    perturbation.
    (Full details of the simulation are
    given in Appendix~\ref{appendix:numerics}.) Successive panels show computations with
    increasingly refined numerical grids, with cell spacing
    $\Delta x = $
     (\emph{a})~$5\times 10^{-4}$m,
     (\emph{b})~$5\times 10^{-5}$m and
     (\emph{c})~$5\times 10^{-6}$m.
     The inlaid panels in~(\emph{a}) and~(\emph{b}) show the corresponding 
     $t = 1\mathrm{s}$ snapshots using shorter vertical axes, as indicated.
     Movie~1 in the Supplementary Material shows an animation of the simulations.
    }
    \label{fig:illp example}%
\end{figure}

Shallow debris flow models with two phases possess a similar mathematical
structure and can suffer from the same pathology.
An illustration of this is depicted in figure~\ref{fig:illp example},
which shows successive attempts to numerically simulate a small
perturbation to a steady uniform flow in the model of~\cite{Meng2022}, 
for conditions where strict hyperbolicity is lost.
While at the coarsest resolution, there appears to be no instability, finer
discretisations reveal oscillations. These develop more rapidly, and with higher
spatial frequency as the grid is refined further. This is because each
successive discretisation permits the approximation of higher wavenumber modes,
thereby inviting faster and faster growth. Any attempt to converge the
simulation towards an underlying solution of the governing equations is
guaranteed to fail, since there is no upper bound on growth rate, implying
that the observed divergence of successive numerical solutions can never
terminate. More precisely, no well defined time-evolving solution
of the continuous equations exists to converge upon. Full details of
this computation are given in Appendix~\ref{appendix:numerics}.

Ill posedness presents a problem for any physical model and numerous examples
have arisen in the fluid mechanics literature over the years~\citep{Joseph1990}.
In particular, it has been discovered to affect mixed-sediment shallow flow
systems that feature particle segregation~\citep{Woodhouse2012,Baker2016} and
bed
morphodynamics~\citep{Cordier2011,Stecca2014,Chavarrias2018,Chavarrias2019,Langham2021}.
Furthermore, it was established long ago that the underlying mixture equations
from which shallow multi-phase debris flow models are derived can feature ill
posedness in some cases~\citep{Bedford1983,Drew1983}.  Though these cases are obviously
physically related, depth-averaged debris flow systems are structurally
inequivalent in general and require their own analyses that depend upon the
particular assumptions employed to reach a shallow model description. There has been
comparatively little work in this direction, possibly because the corresponding
linear dispersion relations (which underlie the analysis of ill posedness) are
at least quartic, making them very difficult to make sense of algebraically.
The only substantial progress appears to be the analysis of~\cite{Pelanti2008},
who derived equations very similar to the model of~\cite{Pitman2005} and
provided inexact bounds on the flow properties that guarantee well posedness.
Nevertheless, these bounds can be violated in situations accessible to realistic
debris flows -- a possibility which should trouble any operational modellers
aiming to compute reliable simulations of these dangerous phenomena.

The following paper presents an investigation of this issue from a general
framework that addresses many of the existing multi-phase models in the
literature. Rather than attempting to identify conditions where models can be
safely used, we instead take the view that any ill posedness within physically
realistic limits is disqualifying for a model and look for situations where this
can occur.
Our analysis is sufficiently general in scope to establish the existence of ill
posedness within the two-phase models of~\cite{Pitman2005}, \cite{Pelanti2008},
\cite{Pudasaini2012}, \cite{Meyrat2022} and \cite{Meng2022}, as well as in the
three-phase model of \cite{Pudasaini2019}.  
The two-phase models are introduced in~\S\ref{sec:theory}, as particular
cases within a generalised shallow-layer description, and their posedness is analysed
in~\S\ref{sec:local analysis}, along with a separate treatment of three-phase models.
Furthermore, in~\S\ref{sec:regularisation} we show that ill
posedness may be alleviated in each of these models via the inclusion of
neglected momentum diffusion terms. En route to this conclusion, a theoretical
recipe is developed for
assessing posedness that may be employed to analyse any model of $n$ phases and
spatial derivatives of up to second order.

\section{Depth-averaged theory}
\label{sec:theory}
Consider a fluid medium consisting of $n$ continuous phases. Each phase $i$
consists of material of constant density $\rho_i$, flows
with velocity $\vect{u}_i \equiv \vect{u}_i(\vect{x},t)$ and occupies
a fraction $\varphi_i\equiv \varphi_i(\vect{x}, t)$ of the mixture volume
at each point in space $\vect{x}$ and time $t$. 
The interior of the flow is assumed to be saturated, so
$\varphi_1+\ldots+\varphi_n=1$. In debris flows, the different phases may
either be fluids, such as pure water or muddy suspensions, or distributions of
small solid particles that are concentrated enough to transmit internal
stresses.
Although no single point may be simultaneously occupied by fluid and particles,
the local volume fractions may be theoretically rationalised either via an explicit assumption that
the phases are everywhere superposed, or by means of suitable averaging
procedures defined over the microscale~\citep{Bedford1983,Jackson2000}.
While there are some technical differences between these
approaches~\citep{Joseph1990b}, the 
resulting form of the governing equations for each phase in three spatial
dimensions is well
established~\citep[see e.g.][]{Anderson1967,Drew1983,Morland1992}.
Assuming negligible surface tension at any interfaces and that no exchange of material
occurs, either between phases or with the external environment,
these may be written as
\begin{subequations}
\begin{gather}
    \frac{\partial \varphi_i}{\partial t}
    + \nabla \cdot (\varphi_i \vect{u}_i)=0,%
    \label{eq:3d eq 1}\\
        \frac{\partial ~}{\partial t}\left(
        \rho_i\varphi_i \vect{u}_i\right)
        + \nabla\cdot
        \left(\rho_i \varphi_i \vect{u}_i \otimes \vect{u}_i \right)
    = \nabla \cdot \matrgr{\sigma}_i + \vect{f}_i - \rho_i \varphi_i \vect{g},
    \label{eq:3d eq 2}%
\end{gather}
\end{subequations}
for $i = 1,\ldots,n$, where $\vect{\sigma}_i$ denotes an
effective (or `partial') stress tensor for each phase,
$\vect{f}_i$ is the total force per unit volume acting on phase $i$ due
to the others and $\vect{g}$ is acceleration due to gravity.

On the grounds that debris flows propagate over distances far greater than their
characteristic thickness, the models that we study simplify
Eqs.~\eqref{eq:3d eq 1} and~\eqref{eq:3d eq 2} by averaging the motion 
over the flow depth.
In addition to this assumption, two simplifications are made for ease of
presentation that do not affect the generality of our primary conclusions.
Firstly, we suppose that the flow propagates over a flat surface located at
$z=0$ 
through which there is no flux of material, 
and orient Cartesian spatial
coordinates $\vect{x} = (x,y,z)$ so that $x$ and $y$ are parallel with this
surface. Secondly, we enforce uniformity of flow in $y$ and hereafter drop 
consideration of this direction from the analysis.  The flow is bounded above
the base by a surface located at $z = h(x,t)$, which is assumed to be stress free.
For any quantity $q(x,z,t)$, its depth-averaged counterpart
$\widebar{q}(x,t)$, is defined by 
\begin{equation}
    \widebar{q}(x, t) = \frac{1}{h}\int_0^h q(x,z,t) \,\mathrm{d}z.
\end{equation}
On averaging 
both sides of Eqs.~\eqref{eq:3d eq 1}
and~\eqref{eq:3d eq 2}, one may obtain
\begin{subequations}
\begin{gather}
    \frac{\partial ~}{\partial t}
    \left(
    h \average{\varphi_i}
    \right) + 
    \frac{\partial ~}{\partial x}
    \left(
    h \average{\varphi_i} \average{u_i}
    \right) = 0,%
    \label{eq:depth av 1}\\
    \frac{\partial ~}{\partial t}\left(\rho_i h \average{\varphi_i}
    \average{u_i}\right)
    + \frac{\partial ~}{\partial x}
    \left(
    \rho_i h \average{\varphi_i} \average{u_i}^2
    - h\widebar{{\sigma}_i^{xx}}
    \right)
    = 
    h \widebar{f_i^x} - \rho_i h \average{\varphi_i} g^x
    -\sigma_i^{xz}|_{z=0},
    \label{eq:depth av 2}%
\end{gather}
\end{subequations}
where algebraic superscripts denote components of vectors and tensors in the
corresponding Cartesian directions.  The details involved in deriving the above
equations follow standard methods and are not important here, except to note
that wherever a product of depth-averaged quantities arises, we make use of the
approximation
\begin{equation}
    \widebar{qr}
    = \widebar{q} \widebar{r}\left[
        1 + \frac{1}{h}\int_0^h 
        \Big( 1 - \frac{q}{\widebar{q}}\Big)
        \left( 1 - \frac{r}{\widebar{r}}\right)
        \mathrm{d}z
    \right]
    \approx \widebar{q}\widebar{r},
    \label{eq:shape factors}%
\end{equation}
where $q$ and $r$ denote arbitrary fields. The relative error introduced by
using this formula is quantified by the second term inside the square brackets
of Eq.~\eqref{eq:shape factors}. It is small if the fields do not vary greatly
over the flow depth. This is frequently assumed in operational models,
including each of the systems that we focus on below.

The framework encapsulated by Eqs.~\eqref{eq:depth av 1} and~\eqref{eq:depth av
2} is general enough to encompass most shallow multi-phase flow models.
Different specialisations to the particular case of debris flows are made by
specifying constitutive models for $\widebar{\sigma_i^{xx}}$, $\widebar{f_i^x}$ and
the basal drag $\sigma_i^{xz}|_{z=0}$. These mostly involve a fluid phase of
either pure water, or water containing fine suspended sediments, and a solids
phase of monodisperse grains. Therefore, for the remainder of this exposition,
we simplify to two phases, labelled $f$ (fluids) and $s$ (solids).
For convenience, a table is provided for this case in
Appendix~\ref{appendix:notation}, which cross-references our notation against
the primary models covered below.
Later, a three-phase model, due to~\cite{Pudasaini2019}, is analysed and its
relevant features are specified separately, in \S\ref{sec:three phase} and
Appendix~\ref{appendix:3 phase}.

One ingredient that must be included within the interphase force terms is
the buoyancy felt by the immersed particles. This is caused by the
the fluid pressure $p$ acting on the solid phase. Therefore, we write the force
on the solids as
\begin{equation}
    \vect{f}_s = -\varphi_s \nabla p + \vect{d}_s,
    \label{eq:fs}%
\end{equation}
where $\vect{d}_s$ represents any other forces associated with the fluid phase
acting on the solids and $p$ is the fluid pressure, which is implied to be
hydrostatic at leading order, by the assumption of shallow flow~\citep[see
e.g.][]{Pitman2005,Meng2022}.  Hydrostatic pressure is determined by the weight
of the fluid within in the water column,
\begin{equation}
    p(z) = \rho_f g^z (h - z).
    \label{eq:hydrostatic}%
\end{equation}
Therefore, on depth-averaging the slope-parallel component of Eq.~\eqref{eq:fs},
we obtain
\begin{equation}
    h\widebar{f^x_s} = 
    -\rho_f g^z 
    h
    \average{\varphi_s}
    \frac{\partial h}{\partial x}
    + h\widebar{d^x_s}.
    \label{eq:depth av buoyancy}%
\end{equation}
By Newton's third law, an equal and opposite force $\widebar{f^x_f} =
-\widebar{f^x_s}$ acts upon the fluid phase.

The remaining component of the interphase forces, $\vect{d}_s = -\vect{d}_f$, must
include contributions due to their relative motion. In conditions close to
equilibrium, this may be modelled with an appropriate closure depending on the relative
velocity $\vect{u}_f - \vect{u}_s$ that captures the
aggregate effect of drag between the two phases~\citep{Morland1992,Jackson2000}.
However, if one phase accelerates into the other, this induces an additional
transfer of momentum between the phases, which can also be
included~\citep{Anderson1967}. The force on individual particles associated with
this is called the `added' or `virtual' mass effect and depends on the relative
accelerations in a frame following the particle~\citep{Maxey1983}.
It is unclear how best to aggregate this into a force acting on a collective
phase of particles, so approaches differ~\citep{Anderson1967,Bedford1983,Drew1983}.  One
option, favoured by~\cite{Pudasaini2012} in the derivation of their debris flow
model, defines the added mass force on the solids to be
\begin{equation}
    \vect{M}_s = 
    C \rho_f \varphi_s \left(
    \frac{\partial \vect{u}_f}{\partial t} + \vect{u}_f \cdot \nabla \vect{u}_f
    -
    \frac{\partial \vect{u}_s}{\partial t} - \vect{u}_s \cdot \nabla \vect{u}_s
    \right)\!,
\end{equation}
where $C$ is a positive coefficient (that may depend on the flow variables, in
particular, the volume fraction).
Depth-averaging this term proceeds in the same way as for the convective terms
on the left-hand side of the governing equations and leads to
\begin{equation}
    h\widebar{M_s^x} = \widebar{C'}\left[
        \frac{\partial~}{\partial t}(\rho_f h\average{\varphi_f}\average{u_f})
        +
        \frac{\partial~}{\partial x}(\rho_f h\average{\varphi_f}\average{u_f}^2)
        \right]
    -
    \gamma\average{C}\left[
        \frac{\partial~}{\partial t}(\rho_s h\average{\varphi_s}\average{u_s})
        +
        \frac{\partial~}{\partial x}(\rho_s h\average{\varphi_s}\average{u_s}^2)
        \right]\!,
    \label{eq:depth av added mass}%
\end{equation}
where $\gamma \equiv \rho_f/\rho_s$ and $C' \equiv C \varphi_s/\varphi_f$.
An opposing force $\widebar{M_f^x} = -\widebar{M_s^x}$ must likewise appear in
the depth-averaged momentum equation for the fluid phase.

The remaining terms to be specified are: the depth-averaged lateral stresses
$\average{\sigma^{xx}_i}$,
the basal stresses $\sigma^{xz}_i|_{z=0}$ and any remaining depth-averaged forces
$h(\widebar{d_i^x}-\widebar{M_i^x})$ (such as drag between the phases, for example).
The choice of the lateral stress components is responsible for most of the key
differences that affect the analysis of models in this paper.
Therefore, these are given with reference to particular models in
the subsections below.
The other two terms will not be given explicitly.
Only terms containing time or space
derivatives of the flow fields affect the analysis in the rest of this paper,
and typically, neither $\sigma^{xz}_i|_{z=0}$, nor $\widebar{d^x_i}$ carry
dependence on gradient information.
Therefore, these are left arbitrary and notation is subsequently simplified by
defining
\begin{equation}
    S_i = (\rho_i h\average{\varphi_i})^{-1}
    \!
    \left[h\left(\widebar{d^x_i} - \widebar{M_i^x}\right) -
    \rho_i h\average{\varphi_i} g^x -
    \sigma_i^{xz}|_{z=0}\right]\!,
\end{equation}
for use in the following subsections. The factor of $1/(\rho_i h
\average{\varphi_i})$ is included to account for the fact that the
momentum
equations will shortly be multiplied through by this quantity in the course of
converting them to quasilinear form.

\subsection{Pitman and Le's model}
\label{sec:pitman le}%
The assumption of shallow flow, used in deriving Eqs.~\eqref{eq:depth av 1}
and~\eqref{eq:depth av 2}, may also be used to infer from the slope-normal
component of Eq.~\eqref{eq:3d eq 2}, that at leading order the normal stresses
are in equilibrium with the interphase forces and gravity
\begin{equation}
    \frac{\partial \sigma_i^{zz}}{\partial z} = -f_i^z + \rho_i \varphi_i g^z.
    \label{eq:z force balance}%
\end{equation}
In deriving their debris flow model, \cite{Pitman2005} use this to obtain
expressions for the stresses. The fluid tensor is assumed to be isotropic and
the slope-normal interphase forces are considered to be dominated by buoyancy,
so $d_s^z = 0$ and from Eq.~\eqref{eq:fs}, $f_s^z = -\varphi_s \partial p /
\partial z = -f_f^z$.  Substituting this into Eq.~\eqref{eq:z force balance},
depth-integrating twice and
using Eq.~\eqref{eq:hydrostatic}, gives
\begin{subequations}
\begin{equation}
    \sigma_f^{zz} = -\rho_f g^z(h-z) \quad\mathrm{and}\quad
    \widebar{\sigma_f^{xx}}=\widebar{\sigma_f^{zz}} = 
    -\frac{1}{2}\rho_f g^z h.
    \tag{\theequation\emph{a,b}}%
\end{equation}
    \label{eq:pitman le fluid stress}%
\end{subequations}
Note that the direction of the buoyancy force and gravity coincide to make the
effective stress for the fluid phase equal to the intrinsic pressure $p$ of the
fluid. Conversely, for the solids phase, buoyancy acts against gravity to reduce
the effective normal stress to
\begin{equation}
    \sigma_s^{zz}= -\varphi_s (\rho_s-\rho_f) g^z (h-z).
    \label{eq:pitman le stress}%
\end{equation}
Since the flow is anticipated to be densely packed with grains,
principles of soil mechanics are invoked to infer a proportional relationship
between lateral and normal stresses, via an Earth pressure coefficient $K$:
\begin{equation}
    \sigma_s^{xx} = K \sigma_s^{zz}.
    \label{eq:earth pressure}%
\end{equation}
On depth-averaging and using Eq.~\eqref{eq:earth
pressure}, one may therefore deduce that
\begin{equation}
    -\frac{\partial ~}{\partial x}\left(
    h\widebar{\sigma_s^{xx}}
    \right)
    =
    \frac{\partial~}{\partial x}\left[
        \frac{1}{2} K (1-\gamma) \rho_s g^z \average{\varphi_s} h^2
    \right]\!.
\label{eq:pitman le pressure grad}%
\end{equation}

The model may be expressed in full by substituting Eqs.~\eqref{eq:depth av
buoyancy}, (\ref{eq:pitman le fluid stress}\emph{b}) and~\eqref{eq:pitman le pressure grad} into
Eqs.~\eqref{eq:depth av 1} and~\eqref{eq:depth av 2} and algebraically simplifying.
It is convenient at this stage, to define variables that express the proportion
of the mixture depth occupied by each phase:
\begin{equation}
    H_i = \average{\varphi_i} h.
    \label{eq:partial depths}%
\end{equation}
Using these variables and noting in particular that $h = H_s + H_f$,
the following set of equations are obtained:
\begin{subequations}
\begin{gather}
    \frac{\partial H_s}{\partial t} + \frac{\partial~}{\partial x}(H_s
    \average{u_s}) =
    0,\label{eq:pitman le 1}\\
    \frac{\partial \average{u_s}}{\partial t} + \average{u_s}\frac{\partial
    \average{u_s}}{\partial x} 
    +
    g^z
    \! \left[
        \gamma + 
        K(1-\gamma)\left(
        1 + \frac{H_f}{2H_s}
        \right)
    \right]
    \! \frac{\partial H_s}{\partial x}
    +
    g^z
    \! \left[
        \gamma + 
        \frac{K}{2} (1 - \gamma)
    \right] 
    \! \frac{\partial H_f}{\partial x}
    = S_s,\label{eq:pitman le 2}\\
    \frac{\partial H_f}{\partial t} + \frac{\partial~}{\partial x}(H_f
    \average{u_f}) =
    0,\label{eq:pitman le 3}\\
    \frac{\partial \average{u_f}}{\partial t} + 
    \average{u_f}\frac{\partial
    \average{u_f}}{\partial x} 
    + g^z \frac{\partial H_s}{\partial x}
    + g^z \frac{\partial H_f}{\partial x}
    = S_f.\label{eq:pitman le 4}
\end{gather}
\label{eq:pitman le}%
\end{subequations}
The particular case of $K = 1$ was later studied in detail
by~\cite{Pelanti2008}.

\subsection{Meng \emph{et al.'s} model}
\label{sec:mengmodel}%
The model of~\cite{Meng2022} is derived using a conceptually different
description of the flow, that posits separate free surfaces for the depth of
solid particles $h_s$ and depth of fluid $h_f$.  When $h_f > h_s$, the particles
are `oversaturated' with fluid and assumed to have settled into a layer at the
bottom of the flow, within which they occupy a constant volume fraction
$\varphi_c$.  We consider this case only, since the analysis of \cite{Meng2024}
(in their Appendix~A) establishes that their model equations in the
`undersaturated' regime $h_f < h_s$ are hyperbolic, with a differential operator
whose structure decouples into separate shallow-layer terms for each phase,
thereby leading to well-posed initial value problems. 

The solids stresses take the same form as in the~\cite{Pitman2005}
model's Eq.~\eqref{eq:pitman le stress}, except they are only present up to the
height $h_s$ of the solids layer,
implying that the term inside
the pressure derivative of Eq.~\eqref{eq:pitman le pressure grad} differs by a
factor of $h_s/h_f$.  Moreover, $K = 1$ is assumed.  Therefore,
\begin{equation}
    -\frac{\partial ~}{\partial x}\left(
    h\widebar{\sigma_s^{xx}}
    \right)
    =
    \frac{\partial~}{\partial x}\left[
        \frac{1}{2} (1-\gamma) \rho_s g^z \average{\varphi_s} h_s h_f
    \right]\!.
    \label{eq:meng solid stress av}%
\end{equation}
Additionally, the viscous component of the fluid stress tensor is retained.
Therefore, rather than appealing to Eq.~\eqref{eq:z force balance}, the
constitutive relation
\begin{equation}
    \matrgr{\sigma}_f = -p \matr{I} + \varphi_f \eta_f \left[\nabla \vect{u}_f +
    (\nabla \vect{u}_f)^T\right]\!,
    \label{eq:meng fluid stress}%
\end{equation}
is proposed, where $\eta_f$ is the dynamic viscosity of the fluid.  The
intrinsic pore fluid pressure is hydrostatic as before, so
Eq.~\eqref{eq:hydrostatic} applies and consequently,
\begin{equation}
    -\frac{\partial~}{\partial x}\left(
    h \widebar{\sigma_f^{xx}}
    \right)
    =
    \rho_f g^z h_f \frac{\partial h_f}{\partial x}
    - \frac{\partial~}{\partial x}\left(
    2 \eta_f h_f \average{\varphi_f} \frac{\partial \average{u_f}}{\partial x}
    \right)\!.
    \label{eq:meng fluid stress av}%
\end{equation}
To obtain the final term on the right, $\average{\partial u_f/\partial x}
\approx \partial\average{u_f}/\partial x$ is used, which follows from an
assumption of low shear in the velocity profile $\average{u_f} \approx u_f(h)$,
and is consistent with the approximation made in Eq.~\eqref{eq:shape factors}.

Averaging the solids volume fraction over the full depth gives
$\average{\varphi_s} = \varphi_c h_s / h_f$. This implies that the equivalent partial
depths [Eq.~\eqref{eq:partial depths}] in this model are
\begin{equation}
    H_s = \varphi_c h_s, \quad H_f = h_f - \varphi_c h_s.
    \label{eq:meng transformations}%
\end{equation}
On making these transformations, the derivative terms in the~\cite{Meng2022}
model equations are the same as the \cite{Pitman2005} 
model's~(\ref{eq:pitman
le}\emph{a--d}), save for the components 
related to the different formulations
for internal stresses.
Therefore, we report only the solid and fluid momentum equations, which may be
obtained by substituting Eqs.~\eqref{eq:meng solid stress av}
and~\eqref{eq:meng fluid stress av}, along with the buoyancy forces [Eq.~\eqref{eq:depth av
buoyancy}], into Eq.~\eqref{eq:depth av 2}, using~\eqref{eq:meng
transformations} and simplifying, leading to
\begin{subequations}
\begin{gather}
    \frac{\partial \average{u_s}}{\partial t} 
    + \average{u_s}\frac{\partial \average{u_s}}{\partial x} 
    + g^z
    \!
    \left[
        \gamma + 
        \frac{1-\gamma}{\varphi_c}
    \right]
    \!
    \frac{\partial H_s}{\partial x}
    +
    g^z\gamma
    \frac{\partial H_f}{\partial x}
    = S_s,\label{eq:meng 1}\\
    \frac{\partial \average{u_f}}{\partial t} + 
    \average{u_f}\frac{\partial
    \average{u_f}}{\partial x} 
    + g^z \frac{\partial H_s}{\partial x}
    + g^z \frac{\partial H_f}{\partial x}
    =
    \frac{2\eta_f}{\rho_f H_f} \frac{\partial~}{\partial x}\left(
    H_f \frac{\partial \average{u_f}}{\partial x}
    \right)
    +
    S_f.
    \label{eq:meng 2}%
\end{gather}
\end{subequations}

A typical choice for the solids fraction constant in the regimes relevant to
this model might be expected to lie somewhere in the range $0.5 \lesssim
\varphi_c \lesssim 0.75$~\citep{Pierson1995}.
Nevertheless, it should be noted that in the limit $\varphi_c \to 1$ (where
there are no saturated gaps between particles) and assuming also that $\eta_f =
0$, Eqs.~\eqref{eq:meng 1} and~\eqref{eq:meng 2} together with~\eqref{eq:pitman
le 1} and~\eqref{eq:pitman le 3} reduce to a system of depth-averaged equations
for the motion of two immiscible fluids of different densities, whose properties
have been widely studied~\citep[see
e.g.][]{Ovysannikov1979,Vreugdenhil1979,Castro2001,Abgrall2009,Kurganov2009,Chiapolino2018}.
A model of this latter type has also been proposed by \cite{Meyrat2022}, for use
in debris flow modelling.

\subsection{Pudasaini's model}
\label{sec:pudasaini}%
\cite{Pudasaini2012} uses an approach that is
consistent with~\cite{Pitman2005}, but extends their framework in various ways. Of
relevance to our analysis are the inclusion of the added mass term given
previously in Eq.~\eqref{eq:depth av added mass} and a fluid stress tensor that
incorporates a non-Newtonian component.

The inclusion of added mass augments the inertial terms in the momentum
equations. 
The coefficient $\average{C}$ in Eq.~\eqref{eq:depth av added mass}
is assumed to be a constant. Furthermore, in order to simplify the conservative
form of the equations~\cite{Pudasaini2012} makes the assumption that
$\average{C'} \equiv
\average{C}\average{\varphi_s}/\average{\varphi_f}$ may be absorbed into the
time and space derivatives of Eq.~\eqref{eq:depth av added mass} without
explicitly holding it constant. This does not appear to be justified in our
view. Nevertheless, summing the added mass force terms for each phase with the
corresponding inertial terms from Eq.~\eqref{eq:depth av 2}
and converting to quasilinear form (i.e.\ by dividing through by $\rho_i H_i$
and simplifying) leads to
\begin{subequations}
\begin{gather}
    (1+\gamma\average{C})\left(
    \frac{\partial \average{u_s}}{\partial t}
    + \average{u_s} \frac{\partial \average{u_s}}{\partial x}
    \right) 
    -\gamma\average{C}\left(
    \frac{\partial \average{u_f}}{\partial t}
    + \average{u_f} \frac{\partial \average{u_f}}{\partial x}
    \right) 
    -\underbrace{\frac{\gamma \average{C}\average{u_f}}{H_s}
    \left[
        \frac{\partial H_{s}}{\partial t}
        + \frac{\partial ~}{\partial x}\left(H_s \average{u_f}\right)
        \right]}_{\mathrm{extra~terms}},
    \label{eq:added mass 1}\\
    \left(1+ \frac{\average{C}H_s}{H_f}\right)
    \left(
    \frac{\partial \average{u_f}}{\partial t} + \average{u_f}\frac{\partial
    \average{u_f}}{\partial x}
    \right)
    -\frac{\average{C}H_s}{H_f}\left(
    \frac{\partial \average{u_s}}{\partial t} + \average{u_s}\frac{\partial
    \average{u_s}}{\partial x}
    \right)
    +\overbrace{\frac{\average{C}\average{u_f}}{H_f}
    \left[
    \frac{\partial H_s}{\partial t} + \frac{\partial~}{\partial
    x}\left(H_s\average{u_f}\right)
    \right]},\label{eq:added mass 2}
\end{gather}
\end{subequations}
for the inertia of the solids and fluid phases respectively.
The extra terms, highlighted by the braces do not appear if Eq.~\eqref{eq:depth
av added mass} is depth averaged directly and could arguably be omitted, since
they correspond to a force between the phases whose physical origin is unclear.
However, in order to analyse the model as it has appeared in prior publications,
we retain them.

The assumed form of the fluid stress tensor is equal to the expression used
by~\cite{Meng2022}, given in Eq.~\eqref{eq:meng fluid stress}, plus an
additional phenomenological component
\begin{equation}
    -\eta_f \mathcal{A}\left[
        \nabla\varphi_s \otimes (\vect{u}_f - \vect{u}_s)
        +
        (\vect{u}_f - \vect{u}_s) \otimes \nabla\varphi_s
        \right]\!,
\end{equation}
where $\mathcal{A}$ is a parameter that depends on the solids
fraction. 
After adding on the Newtonian component, depth averaging $\sigma_f^{xx}$ gives
\begin{equation}
    -\frac{\partial~}{\partial x}\left(
    h \widebar{\sigma_f^{xx}}
    \right)
    =
    \rho_f g^z h \frac{\partial h}{\partial x}
    - \frac{\partial~}{\partial x}\left(
    2 \eta_f h \average{\varphi_f} \frac{\partial \average{u_f}}{\partial x}
    - 2\eta_f \average{{\mathcal{A}}} h
    (\average{u_f}-\average{u_s})\frac{\partial \average{\varphi_s}}{\partial x}
    \right)
    \!,
    \label{eq:pudasaini fluid stress av}%
\end{equation}
for this model, where we have used $\average{\partial \varphi_s/\partial x} \approx \partial
\average{\varphi_s}/\partial x$, which is consistent with the assumption of
negligible variation in volume fraction over the depth, $\average{\varphi_s}
\approx \varphi_s(h)$. In the original derivation, \cite{Pudasaini2012} goes
further, following an approach of \cite{Iverson2001} for averaging diffusive
stresses by bringing $h$ outside the spatial derivatives of
Eq.~\eqref{eq:pudasaini fluid stress av}. This introduces extra terms, which,
under the stress-free boundary condition reduce to expressions that do not
contain derivatives and may be modelled separately as source
terms~\citep{Pudasaini2012}. These extra steps do not affect the forthcoming
analysis of the model structure (since the linearised diffusion operator remains
the same). Therefore, we leave Eq.~\eqref{eq:pudasaini fluid stress av} as it
is.

\section{Local analysis}
\label{sec:local analysis}%
We will demonstrate that the two-phase models outlined in the previous section,
as well as straightforward three-phase extensions to these systems, contain flow
regimes where the equations are
ill posed as initial value problems.  This is because under certain
conditions, infinitesimal disturbances blow-up with linear growth rates
that increase without bound in the limit of high spatial frequencies, leaving
the equations without solutions -- a pathological property sometimes known as a
`Hadamard instability'~\citep{Joseph1990}.

\subsection{Two-phase models}
\label{sec:two-phase}%
Given some putative model solution with fields
$\vect{q} = (H_s, \average{u_s}, H_f, \average{u_f})^T$\!,
we would like to understand the
local behaviour of the governing equations at an arbitrary space-time 
location $(x_0,t_0)$. 
Denote a state vector there, by
\begin{equation}
    \vect{q}_0 = \vect{q}(x_0, t_0) = \left(
    H_{s}^{(0)}\!\!,~\widebar{u_s}^{(0)}\!,~H_{f}^{(0)}\!\!,~\widebar{u_f}^{(0)}
    \right)^T\!\!\!.
\end{equation}
We assume non-vanishing fluid depth $H_f^{(0)} > 0$ and velocity
$\widebar{u_f}^{(0)} \neq 0$, so that the governing equations for each model may
be non-dimensionalised with respect to these scales. 
States may then be fully characterised by three dimensionless quantities:
\begin{subequations}
\begin{equation}
    R_H = H_s^{(0)} / H_f^{(0)}, \quad R_u = \widebar{u_s}^{(0)} /
    \widebar{u_f}^{(0)}, \quad
    \Fr = \frac{\widebar{u_f}^{(0)}}{\sqrt{g^z H_f^{(0)}}},
    \tag{\theequation\emph{a--c}}%
\end{equation}
\end{subequations}
where $\Fr$ is the local Froude number for the fluid phase.
Therefore, hereafter the transformations
\begin{subequations}
\begin{equation}
    x\mapsto x / H_f^{(0)}\!, ~
    t \mapsto t \widebar{u_f}^{(0)} / H_f^{(0)}\!, ~
    H_i \mapsto H_i / H_f^{(0)}\!, ~
    \average{u_i} \mapsto \average{u_i} / \widebar{u_f}^{(0)}\!, ~
    S_i \mapsto S_i H_f^{(0)} / (\widebar{u_f}^{(0)})^2,
    \tag{\theequation\emph{a--e}}%
    \label{eq:nondimensionalisation}%
\end{equation}
\end{subequations}
are made
to the two-phase models analysed.
Furthermore, note that the systems detailed in~\S\ref{sec:pitman
le}--\ref{sec:pudasaini} may be collectively cast in the
general form
\begin{equation}
    \matr{A}(\vect{q})
    \frac{\partial \vect{q}}{\partial t}
    + \matr{B}(\vect{q}) \frac{\partial \vect{q}}{\partial x}
    = \vect{S}(\vect{q})
    + \matr{D}_1(\vect{q})\frac{\partial~}{\partial x}\left(
    \matr{D}_2(\vect{q})
    \frac{\partial \vect{q}}{\partial x}
    \right),
    \label{eq:gen model eqs}%
\end{equation}
where 
$\vect{S}(\vect{q}) = (0, S_s, 0, S_f)^T$
and $\matr{A}$, $\matr{B}$, $\matr{D}_1$, $\matr{D}_2$
are matrices of (dimensionless) variable coefficients 
that that may be readily specified for each model.

We now `freeze' the solution, by assuming that $\vect{q}(x,t) = \vect{q}_0$
within a local neighbourhood of $(x_0, t_0)$ and consider the evolution of a
normal-mode perturbation $\vect{r} \exp(\im k x + \sigma t)$ to this state,
where $k$ is a real-valued wavenumber, $\sigma$ a complex growth rate and
$\vect{r}$ a vector constant with $|\vect{r}| \ll |\vect{q}_0|$.  Linearising
Eq.~\eqref{eq:gen model eqs} around the frozen base state $\vect{q}_0$ leads to
the following eigenproblem for the pair $(\sigma, \vect{r}$):
\begin{gather}
    \sigma \matr{A}(\vect{q}_0)\vect{r} + \im k \matr{B}(\vect{q}_0) \vect{r}
    = \matr{C}(\vect{q}_0)\vect{r} - k^2 \matr{D}(\vect{q}_0) \vect{r},
    \label{eq:full eigenproblem}%
\end{gather}
where $\matr{C} \equiv \partial \vect{S} / \partial \vect{q}$ and $\matr{D} =
\matr{D}_1 \matr{D}_2$.  If $\vect{q}_0$
happens to represent a state for which the model equations admit steady uniform
flow, i.e.\ $\vect{S}(\vect{q}_0)=\vect{0}$, then the solutions to
Eq.~\eqref{eq:full eigenproblem} dictate the linear stability of such a flow,
for which $\Real(\sigma) > 0$ indicates an unstable mode and the 
case of Hadamard instability occurs if $\Real(\sigma) \to \infty$ as $k\to
\infty$, indicating that the governing equations~\eqref{eq:gen model eqs} are
ill posed at $\vect{q}_0$.
Otherwise, the procedure of freezing the base state is justified insofar as it
may be used to identify this latter pathology on the grounds that any candidate
solution $\vect{q}$ must be effectively constant near $(x_0,t_0)$, when measured
with respect to the infinitesimal length and time scales over which the
Hadamard instability develops~\citep{Joseph1990,Joseph2013}.

Since diffusion can be small, relative to other terms in the equations, it is
sometimes desirable to neglect its effects. Therefore, we first consider
the case where $\matr{D}$ is the zero matrix. It may then straightforwardly be
determined from Eq.~\eqref{eq:full eigenproblem} that in the asymptotic limit of
high $k$, the leading order components of the growth rates for the four
eigenmodes are $\sigma = -\im k \lambda_j$, where $\lambda_j$ ($j=1,\ldots,4$) denote the
characteristic wave speeds of the problem, given by the solutions to the
generalised eigenproblem $\matr{B}\vect{r} = \lambda_j \matr{A}\vect{r}$.
If all four are real and distinct then the system is said to be `strictly
hyperbolic' at $\vect{q}_0$ and is well posed as an initial-value problem.  On
the other hand, any complex characteristics must arise in conjugate pairs.
Since one of the pair must have $\Imag(\lambda_j)>0$, the corresponding real
part of $\sigma$ is positive and scales as $O(k)$ for $k \gg 1$, giving rise to
a Hadamard instability.
Repeated
real characteristics can also lead to growth rate blow-up, but the reasons for
this are more subtle.  This case is covered later, in~\S\ref{sec:general
framework}.

\subsubsection{Emergence of ill posedness}
\label{sec:emergence}%
The inclusion of added mass leads to complications, which we address shortly, in
\S\ref{sec:added mass} and~\S\ref{sec:pudasaini diffusive}. If it is neglected,
then $\matr{A}$ simplifies to the identity matrix $\matr{I}$ and the problem
reduces to computing the eigenvalues of the Jacobian $\matr{B}$, which has the
same essential form for each of the models. Bearing in mind our transformation
to dimensionless variables in Eqs.~\eqref{eq:nondimensionalisation}, this matrix
is
\begin{gather}
    \matr{B}(\vect{q}) = \begin{pmatrix}
        \average{u_s} & H_s & 0 & 0 \\
        (\gamma + \beta_1)\Fr^{-2} &
        \average{u_s} & (\gamma + \beta_2)\Fr^{-2} &
        0 \\
        0 & 0 & \average{u_f} & H_f \\
        \Fr^{-2} & 0 & \Fr^{-2} & \average{u_f}
    \end{pmatrix}
    ,
    \label{eq:jacobian}%
\end{gather}
where $\beta_1 = K(1-\gamma)[1 + H_f/(2H_s)]$, $\beta_2 = K(1-\gamma)/2$
for~\cite{Pitman2005} and~\cite{Pudasaini2012}; $\beta_1 =
(1-\gamma)/\varphi_c$, $\beta_2 = 0$ in~\cite{Meng2022}; and $\beta_1 =
1-\gamma$, $\beta_2 = 0$ for two-fluid models~\citep[e.g.][]{Ovysannikov1979}, as well as the debris flow model of
\cite{Meyrat2022}. Note that at
$\vect{q}=\vect{q}_0$, $H_s = R_H$, $\average{u_s} = R_u$ and $H_f =
\average{u_f} = 1$.

The possibility for $\matr{B}(\vect{q}_0)$ to have complex characteristics arises due to
the coupling between the momentum equations provided by the entries $\mathsfi{B}_{23}$
and $\mathsfi{B}_{41}$. Physically, these terms arise because the buoyancy and solids
stresses depend on the total depth $H_s + H_f$. 
For systems without this coupling, i.e.\
$\mathsfi{B}_{23} = \mathsfi{B}_{41}=0$, 
the eigenvalues of $\matr{B}(\vect{q}_0)$ are
\begin{subequations}
\begin{equation}
    \lambda_s^\pm \equiv R_u \pm \frac{\sqrt{R_H(\gamma + \beta_1)}}{\Fr}
    \quad\mathrm{and}\quad
    \lambda_f^\pm \equiv 1 \pm \frac{1}{\Fr}.
    \tag{\theequation\emph{a,b}}%
    \label{eq:uncoupled characteristics}%
\end{equation}
\end{subequations}
These are real provided $\beta_1 + \gamma > 0$. For the models closures described above,
this is certainly the case, since both $\beta_1$ and $\gamma$ are strictly
positive.  While the corresponding expressions for the eigenvalues of
$\matr{B}(\vect{q}_0)$
in the general case, $\mathsfi{B}_{23},\mathsfi{B}_{41}\neq 0$, can be computed via the quartic formula, these are are too
complicated to be especially
useful~\citep{Pitman2005,Pelanti2008,Pudasaini2012}.
Nevertheless, since $\matr{B}(\vect{q}_0)$ is almost block diagonal, its characteristic
polynomial is amenable to further analysis. 

\begin{figure}
    \centering
    \includegraphics[width=\textwidth]{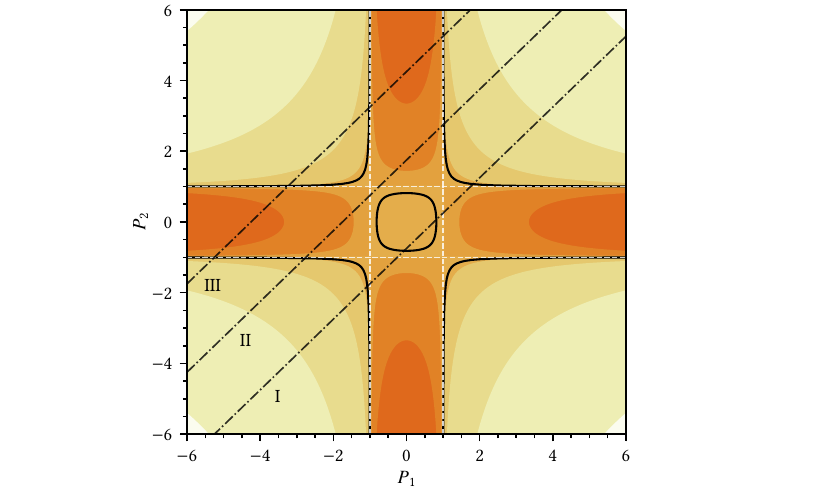}%
    \caption{Geometric analysis of the characteristics for two-phase models.
    Filled contours of the surface $f(P_1,P_2)$ are plotted, spaced
    at intervals $\pm 10^m$ for $m = 0, \ldots, 4$. The zero contour is marked
    separately (white dashed), as is the level set at $c = 1/3$ (black solid). 
    Dash--dotted lines are $P_2 = P_1 - 0.75 + 2n$, for $n = 0,1,2$.
    }
    \label{fig:pq plot}%
\end{figure}

In particular, one can generalise an
approach followed by~\cite{Ovysannikov1979} for the simpler two-fluid case
($\beta_1 = 1-\gamma$, $\beta_2 = 0$) and notice that the eigenvalues $\lambda_j$ 
are determined by an equation of the form
\begin{equation}
    f(P_1, P_2) \equiv (P_1^2-1)(P_2^2-1) = c,
    \label{eq:level set func}%
\end{equation}
where
\begin{subequations}
\begin{gather}
    P_1^2 = \frac{(\lambda_j - R_u)^2\Fr^2}{R_H(\gamma + \beta_1)},\quad
    P_2^2 = (\lambda_j - 1)^2\Fr^2,\quad
    \mathrm{and}\quad
    c = \frac{\gamma + \beta_2}{\gamma + \beta_1}.
    \tag{\theequation\emph{a--c}}%
\end{gather}
    \label{eq:PQK}%
\end{subequations}
For a particular point in parameter space, characterised by the triple $(R_H,
R_u, \Fr)$, we can eliminate $\lambda_j$ from Eqs.~(\ref{eq:PQK}\emph{a,b}) to
determine that the characteristics lie on the intersection of the line
\begin{equation}
    P_2 = P_1\sqrt{R_H(\gamma + \beta_1)} + \Fr(R_u - 1)
    \label{eq:PQ line}%
\end{equation}
with the level set given by the contour of the surface $f(P_1,P_2)$
[Eq.~\eqref{eq:level set func}] at the value~$c$. This is depicted
graphically in figure~\ref{fig:pq plot}(\emph{a}).
Coloured contours in the figure show the surface $f(P_1,P_2)$, with an example
level at $c = 1/3$ given by the solid black curves.
Three dash--dot black lines illustrate
possibilities for the characteristics. The line labelled~I
represents a strictly hyperbolic case, since it possesses four distinct
intersections with the solid black contour. On shifting the line upwards to~II
[by increasing $\Fr(R_u-1)$] the characteristics associated with
the central contour merge to form a complex conjugate pair and only two real
solutions to Eqs.~\eqref{eq:level set func} and~\eqref{eq:PQ line} remain.
Shifting the line further up, recovers strict hyperbolicity, since at
position~III, it makes two additional intersections with the portion of the
level set that is confined to $\{(P_1, P_2) : P_1 < -1, P_2 > 1\}$.  Provided
that $c>0$ and that $\beta_1$, $\beta_2$ are either constants or a functions of $R_H$
only, as is the case for the models considered herein,
we can see that there will always be an ill-posed region associated with the
loss of strict hyperbolicity (i.e.\ regions without four distinct real
eigenvalues).
This is because a given $R_H$ fixes the level set determined by~$c$. Then,
varying $\Fr(R_u - 1)$ shifts the dash--dotted lines in the $P_2$ direction, guaranteeing that
they pass through a region with only two intersections. Indeed, by symmetry,
there must be two such regions.

This framework encapsulates the analysis by~\cite{Pitman2005} who showed for
their model, that cases close to $R_u = 1$ are always strictly hyperbolic.
This is a consequence of the fact that the Eq.~\eqref{eq:PQ line} lines pass
through the origin at this point.
Moreover,~\cite{Pelanti2008} later gave bounds on $|R_u - 1|$ that guarantee
well posedness for sufficiently small and sufficiently large values.

The white dashed lines in figure~\ref{fig:pq plot} show the level set contours at
$c=0$, given by $P_1 = \pm 1$, $P_2 = \pm 1$.  In this special case, the
characteristics are everywhere real and it is straightforward to see that they
must be the same as the values for uncoupled systems, given in
Eqs.~\eqref{eq:uncoupled characteristics}. 
Moreover, each point $(\pm 1, \pm 1)$ may be linked to one of the four possible
intersections between the solid ($\lambda_s^\pm$) and fluid ($\lambda_f^\pm$) 
characteristics.  For example, let
$R_u$, $R_H$ be fixed and suppose that $R_u > 1$, implying that the dash--dotted
lines of figure~\ref{fig:pq plot} [Eq.~\eqref{eq:PQ line}] intercept the
$P_2$-axis at positive values.  From examination of the expressions for the
characteristics in Eqs.~\eqref{eq:uncoupled characteristics}, it may determined
that there always exists a $\Fr$ such that $\lambda_s^- = \lambda_f^+$ in this
case.  Moreover, depending on whether the gradients of the dash--dot lines
$\sqrt{R_H(\gamma+\beta_1)}$ are greater or less than unity, the respective
intersections $\lambda_s^-=\lambda_f^-$ and $\lambda_s^+ = \lambda_f^+$, are
possible.  By considering (geometrically) the corresponding options for the
lines to pass through $(\pm 1, \pm 1)$ in this case, we infer that $\lambda_s^-
= \lambda_f^+$ corresponds to the point $(-1, 1)$ and likewise, that
$\lambda_s^- = \lambda_f^-$ corresponds to $(-1, -1)$ and
$\lambda_s^+=\lambda_f^+$ to $(1,1)$.  Symmetric reasoning for the case $R_u <
1$ determines the final intersection, $\lambda_s^+ = \lambda_f^-$ at $(1, -1)$.
The important points are $(-1, 1)$ and $(1, -1)$, when the positive and negative
branches coincide. This occurs when
\begin{equation}
    R_u = 1 \pm \frac{1}{\Fr}\left[
    1 + \sqrt{R_H(\gamma + \beta_1)}
    \right]\!.
\label{eq:uncoupled intersection}%
\end{equation}
It is from these intersections that the complex eigenvalues of
$\matr{B}(\vect{q}_0)$
emerge when the system is fully coupled.  Therefore, the consequent blow-up in
growth rate in these regions can be thought of as stemming from a resonant
interaction between the characteristic wave speeds of the solid and fluid
phases.

\begin{figure}
    \centering
    \includegraphics[width=\textwidth]{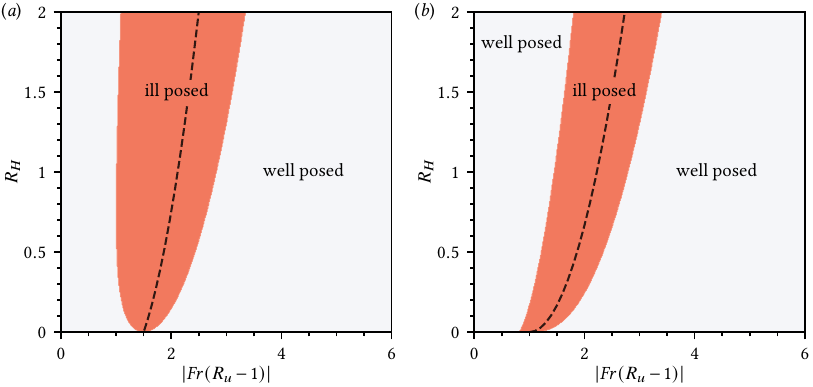}%
    \caption{%
    Regions of parameter space which contain complex characteristics,
    indicated by the red shaded regions, for the models of
    (\emph{a})~\cite{Pitman2005} with $\gamma = 0.5$, $K=1$ and
    (\emph{b})~\cite{Meng2022} with
    $\gamma = 0.5$, $\varphi_c = 0.5$. In the case of~(\emph{b}), the 
    parameter choices correspond to
    the solid black level set in figure~\ref{fig:pq plot}.
    Outside the shaded regions, the characteristics are real and distinct.
    The black dashed curves are where positive and negative branches of
    characteristics from the corresponding uncoupled problems intersect, 
    as given in Eq.~\eqref{eq:uncoupled intersection}.
    }
    \label{fig:twophase posedness}%
\end{figure}

In figures~\ref{fig:twophase posedness}(\emph{a}) and~(\emph{b}), we plot
the regions where ill posedness occurs for the \cite{Pitman2005} and
\cite{Meng2022} models respectively (without diffusion), in terms of $R_H$
and $|\Fr(R_u-1)|$. 
Note that these two parameters fully determine whether the characteristics are
real-valued or not.  As already inferred from the geometric analysis, the models
are unconditionally well posed when $R_u = 1$ and at sufficiently high values of
$|\Fr(R_u-1)|$.  Furthermore, the bands of ill posedness are organised around
the condition in Eq.~\eqref{eq:uncoupled intersection} (black dashed lines).
The width of the bands is contingent on the model parameters, which select the
level set(s) in figure~\ref{fig:pq plot} and the qualitative differences in shape
between the bands for the two models are explained by the different dependence.
Specifically, while the level set value $c$ for the \cite{Meng2022} model is
fixed, for \cite{Pitman2005}, $c\equiv c(R_H)$ with $c\to 0$ as $R_H \to 0$.
This implies that the width of the figure~\ref{fig:twophase posedness}(\emph{a})
band approaches zero in this limit. Conversely, when $c$ is constant, the band
has a finite width as $R_H\to 0$, determined by the minimum distance between the
central piece of the figure~\ref{fig:pq plot} level set and any of the lines $P_1,
P_2 = \pm 1$, which are asymptotically approached by the other sections of the
level set.  A brief calculation shows that this is $1-\sqrt{1-c}$ for $c \in [0,
1]$, or $1$ for $c > 1$ (where in this latter case there is no central piece of
the level set).  Additionally, the uncoupled characteristics intersections must
lie at the upper limit of the band as $R_H \to 0$.  Combining these observations
with Eq.~\eqref{eq:uncoupled intersection} determines that the interval
$(\sqrt{1-c}, 1)$ remains ill posed in this case, as $R_H \to 0$. This property
is demonstrated for the~\cite{Meng2022} model with $c = 1/3$, by examining
figure~\ref{fig:twophase posedness}(\emph{b}).

\subsubsection{Added mass effect}
\label{sec:added mass}%
When the added mass effect is included in the two-phase model of
\cite{Pudasaini2012}, many additional terms are introduced that cause the
equations to be more strongly coupled. Though this model also contains diffusive
terms, it is informative to investigate first how the incorporation of this
additional physics affects the model's eigenstructure in the absence of
diffusion.  One reason for this is that, at least in some cases, ill posedness
in some non-depth-averaged two-phase flow systems without diffusion can be
regularised by including added mass terms~\citep{Drew1983}.

On introducing added mass effects by generalising the inertial terms in the
solids and fluid momentum equations to the
expressions given previously in Eqs.~\eqref{eq:added mass 1}
and~\eqref{eq:added mass 2} respectively,
the matrices $\matr{A}$ and $\matr{B}$ become
\begin{gather}
    \matr{A} = \begin{pmatrix}
        1 & 0 & 0 & 0 \\
        -\gamma \average{C}\average{u_f}H_s^{-1} & 1+\gamma\average{C} & 0 &
        -\gamma\average{C} \\
        0 & 0 & 1 & 0 \\
        \average{C}\average{u_f}H_f^{-1} & -\average{C}H_sH_f^{-1} & 0 &
        1+\average{C}H_sH_f^{-1}
    \end{pmatrix},\label{eq:pudasaini A}\\
    \matr{B} = \begin{pmatrix}
        \average{u_s} & H_s & 0 & 0 \\
        (\gamma +
        \beta_1)\Fr^{-2}-\frac{\gamma\average{C}\average{u_f}^2}{H_s} &
        (1+\gamma\average{C})\average{u_s} & (\gamma + \beta_2)\Fr^{-2} &
        -2\gamma\average{C}\average{u_f} \\
        0 & 0 & \average{u_f} & H_f \\
        \Fr^{-2}+\average{C}\average{u_f}^2H_f^{-1} &
        -\average{C}\average{u_s}H_sH_f^{-1} & \Fr^{-2} & \average{u_f} + 
        \frac{2 \average{C}\average{u_f}H_s}{H_f}
    \end{pmatrix}
    .
    \label{eq:pudasaini B}%
\end{gather}
When the added mass coefficient $\average{C}$ is nonzero, the corresponding
characteristic polynomial $p_c(\lambda) = \det(\matr{B}-\lambda\matr{A})$ for this
system lacks the advantageous structure that was leveraged in the previous
section to analyse the eigenvalues geometrically.  Nevertheless, they are
straightforward to obtain numerically at any point in parameter space. On doing
so, it was found (as might well be expected) that the boundaries of the well-posed
regions do not collapse neatly onto curves in terms of the parameters $R_H$ and
$\Fr(R_u-1)$, as before. However, it is possible to observe the qualitative
effect of increasing $\average{C}$ from zero.

The plots in figure~\ref{fig:pudasaini
posedness} show an illustrative example, in which $\gamma = 0.5$, $R_H = 1$ and
$\average{C}$ is incremented up to the value of $0.5$ suggested
by~\cite{Pudasaini2012}.
\begin{figure}
    \centering
    \includegraphics[width=\textwidth]{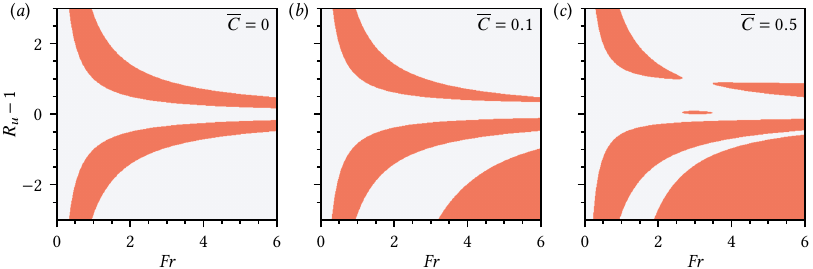}%
    \caption{Effect of added mass terms in the~\cite{Pudasaini2012} model
    without diffusion.
    Regions of parameter space that possess complex characteristics are
    shaded red, for $R_H = 1$, $\gamma = 0.5$ and $\average{C} = $
    (\emph{a})~$0$, (\emph{b})~$0.1$ and~(\emph{c})~$0.5$.
    }
    \label{fig:pudasaini posedness}%
\end{figure}
When $\average{C}=0$, the system reduces to the structure of
the~\cite{Pitman2005} model. The ill-posed regions lie either side of
$R_u = 1$ and take the form of bands around the curves given previously in
Eq.~\eqref{eq:uncoupled intersection}. Increasing $\average{C}$ to $0.1$,
results in a slight narrowing of the `upper' band with $R_u > 1$ and a slight
thickening of the lower $R_u < 1$ band. Additionally, a new region of complex
characteristics emerges beneath the lower band at higher Froude numbers.  This
region extends further towards lower $\Fr$ when $\average{C}$ is increased to
$0.5$ [figure~\ref{fig:pudasaini posedness}(\emph{c})], leaving most of the $R_u <
1$ half-plane ill posed. Furthermore, the upper band separates
into two pieces, leaving a well-posed region in between them. Moreover, a small
region of ill posedness appears at an $R_u$ closer to unity. It is approximately
centred around the point  $(\Fr, R_u) = (3.04,1.05)$.
We inspected
equivalent plots for other choices of $\gamma$ and $R_H$ in the ranges
$0.3<\gamma<0.8$, $0.2<R_H<1.5$ and found them to be qualitatively similar.

These results indicate that the added mass force in this case does little to
ameliorate the problem of ill posedness on its own and arguably seems to make
matters worse, especially when the fluids velocity greatly exceeds the solids
velocity ($R_u<1$) -- a situation which could be encountered when a less
concentrated debris flow entrains a static pile of grains, for example.
Some analytical insight into the emergence of the large ill-posed region for
$R_u < 1$ is gained later in~\S\ref{sec:pudasaini diffusive}.

\subsection{Three-phase models}%
\label{sec:three phase}%
For three-phase models that share the same essential structure as the
two-phase models we have analysed, it is possible to
generalise the geometric reasoning of \S\ref{sec:emergence} to 
identify regions of parameter space that must contain complex characteristics,
provided the added mass effect is negligible.

Therefore, we return to the case $\average{C}=0$, $\matr{A}=\matr{I}$ and
consider models that possess a Jacobian of the form
\begin{equation}
\matr{B} = \begin{pmatrix}
    \average{u_1} & H_1 & 0 & 0 & 0 & 0 \\
    \beta_{11} & \average{u_1} & \beta_{12} & 0 & \beta_{13} & 0 \\
    0 & 0 & \average{u_2} & H_2 & 0 & 0 \\
    \beta_{21} & 0 & \beta_{22} & \average{u_2} & \beta_{23} & 0 \\
    0 & 0 & 0 & 0 & \average{u_3} & H_3 \\
    \beta_{31} & 0 & \beta_{32} & 0 & \beta_{33} & \average{u_3}
\end{pmatrix},
    \label{eq:3 phase jacobian}%
\end{equation}
where $H_i$ denote partial heights for each phase, $\average{u_i}$ the
corresponding downstream
velocities and
$\beta_{ij}$ represent arbitrary functions of these flow variables.
This generalises the essential structure of the two-phase Jacobian in
Eq.~\eqref{eq:jacobian} to three phases.
Denote the characteristic polynomial of this matrix
by $p_c$. By direct computation, it may be shown that $p_c(\lambda) = 0$ simplifies to
\begin{equation}
    f(P_1,P_2,P_3)\equiv\prod_{i=1}^3 (P_i^2-1) - 
        (P_1^2-1)\frac{\beta_{23}\beta_{32}}{\beta_{22}\beta_{33}} -
        (P_2^2-1)\frac{\beta_{13}\beta_{31}}{\beta_{11}\beta_{33}} -
        (P_3^2-1)\frac{\beta_{12}\beta_{21}}{\beta_{11}\beta_{22}}
        =
        c,
        \label{eq:3 phase f}%
\end{equation}
where 
\begin{equation}
    P^2_i = \frac{(\average{u_i} - \lambda)^2}{\beta_{ii} H_i}
\quad    \mathrm{and} \quad c=
        \frac{\beta_{13}\beta_{21}\beta_{32} +
        \beta_{12}\beta_{23}\beta_{31}}{\beta_{11}\beta_{22}\beta_{33}}>0,
        \label{eq:3 phase Pc}%
\end{equation}
for $i = 1,2,3$.  We retain the convention adopted previously, by
non-dimensionalising with respect to the depth and velocity of the third phase,
represented by the final two governing equations and hereafter assumed to
represent the carrier fluid. There are now two pairs of relevant dimensionless
quantities associated with the relative heights and velocities of the phases:
\begin{subequations}
\begin{equation}
    R_{H_1} = H_1^{(0)} / H_3^{(0)}, \quad
    R_{H_2} = H_2^{(0)} / H_3^{(0)}, \quad
    R_{u_1} = \widebar{u_1}^{(0)} / \widebar{u_3}^{(0)}, \quad
    R_{u_2} = \widebar{u_2}^{(0)} / \widebar{u_3}^{(0)},
    \tag{\theequation\emph{a--d}}%
    \label{eq:3 phase nondim}%
\end{equation}
\end{subequations}
alongside the Froude number for the fluid phase $\Fr =
\widebar{u_3}^{(0)}/\sqrt{g^z H_3^{(0)}}$.
On making the appropriate non-dimensionalising transformations 
and eliminating $\lambda$ from among the defining relations for the $P_i$
coordinates in Eq.~\eqref{eq:3 phase Pc}, it may be concluded that the number of
real roots of $p_c$ at a given point in parameter space is determined by the
intersections of the level surface defined in Eqs.~\eqref{eq:3 phase f}
and~\eqref{eq:3 phase Pc}, with the line given by the map
\begin{equation}
    P_3 \mapsto
    \left(
    \frac{R_{u_1}-1}{\sqrt{\beta_{11} R_{H_1}}} + P_3\sqrt{\frac{\beta_{33}
    }{\beta_{11} R_{H_1}}},
    \frac{R_{u_2}-1}{\sqrt{\beta_{22} R_{H_2}}} + P_3\sqrt{\frac{\beta_{33}
    }{\beta_{22} R_{H_2}}},
    P_3
    \right)\!.
    \label{eq:pqr line}%
\end{equation}

To illustrate the resulting geometric picture, we use the model
of~\cite{Pudasaini2019}, which extends the two-phase system
of~\cite{Pudasaini2012} to incorporate an intermediate fraction of fine solid
particles.
When added mass effects are neglected, the Jacobian for this model matches the
structure given in Eq.~\eqref{eq:3 phase jacobian}. If the equations are
organised such that the first two rows denote the solid phase, the second two
the fine-solid phase and the final two the fluid phase, then the 
(non-dimensionalised) $\beta_{ij}$
closure terms are
\begin{subequations}
\begin{equation}
    \beta_{11} = \frac{1}{\Fr^2}\left[
        1 + \frac{1}{2}(1-\gamma_1)\left(
        \frac{R_{H_1} + R_{H_2} + 1}{R_{H_1}}
        \right)
        \right], \quad \beta_{12} = \beta_{13} = \frac{1}{2\Fr^2}(1+\gamma_1)
    \tag{\theequation\emph{a,b}}%
    \label{eq:3 phase jac closures}%
\end{equation}
\end{subequations}
and $\beta_{2i} = \gamma_2/\Fr^2$, $\beta_{3i} = 1/\Fr^2$, for $i = 1,2,3$,
where $\gamma_1$ is the ratio of fluid to solid densities and $\gamma_2$ is the
ratio of fluid to fine solid densities.  These latter two parameters are fixed
material constants. 
On substituting the expressions for $\beta_{ij}$ into
Eqs.~\eqref{eq:3 phase f},~\eqref{eq:3 phase Pc}
and~\eqref{eq:pqr line}, it may be deduced that both the level surface
$f(P_1,P_2,P_3)=c$ and the
gradient of the line in Eq.~\eqref{eq:pqr line} depend only on the flow via the
relative heights $R_{H_1}$ and $R_{H_2}$.
Consequently, for a given $(R_{H_1}, R_{H_2})$ pair, the number of intersections
between the line and the level set is determined by the remaining degrees of
freedom for the line, namely the terms
\begin{subequations}
\begin{equation}
    K_1 \equiv \frac{R_{u_1} - 1}{\sqrt{\beta_{11}R_{H_1}}} 
    \quad\mathrm{and}\quad
    K_2 \equiv \frac{R_{u_2} - 1}{\sqrt{\beta_{22}R_{H_2}}}.
    \tag{\theequation\emph{a,b}}%
\end{equation}
\end{subequations}
By substituting in the appropriate values for $\beta_{11}$, $\beta_{22}$, it may
be seen that $K_i \propto \Fr(R_{u_i} - 1)$.

\begin{figure}
    \includegraphics[width=\textwidth]{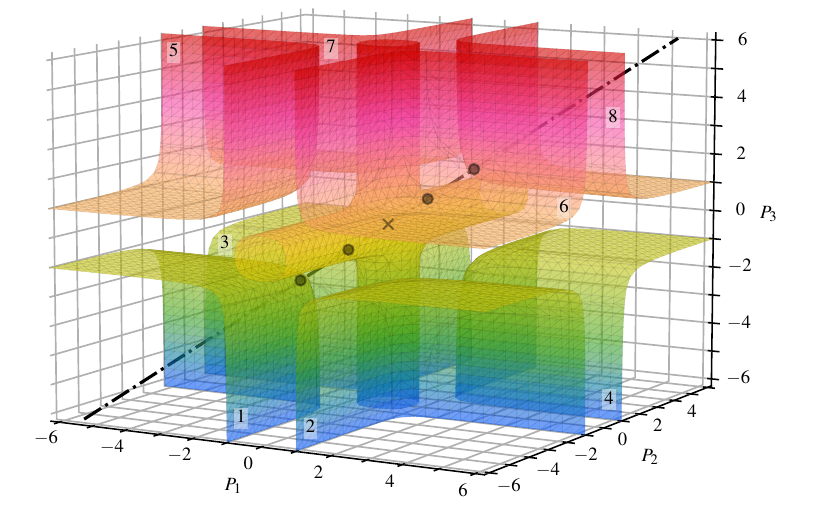}%
    \caption{The surface $f(P_1,P_2,P_3) = c$, for the three-phase model
    of~\cite{Pudasaini2019}, with $\gamma_1 = \gamma_2 = 0.5$ and $R_{H_1} =
    R_{H_2} = 1$. For visual clarity, the disjoint pieces of the surface are
    rendered with a triangular mesh and coloured from blue to red according to
    the value of the $P_3$ coordinate.  Also plotted is the line defined by
    Eq.~\eqref{eq:pqr line} for $R_{u_1}=R_{u_2}=1$. This intersects with the
    surface at the four points marked with circles and at the origin (marked
    with a cross), which is an additional isolated solution of
    $f(P_1,P_2,P_3)=c$, in this case.
    Movie~2 in the Supplementary Material shows an animated view of the surface.
    }
    \label{fig:pqr}%
\end{figure}

In figure~\ref{fig:pqr}, we plot the surface corresponding to the case $R_{H_1} =
R_{H_2} = 1$ and $\gamma_1 = \gamma_2 = 0.5$. 
It consists of nine disjoint pieces, comprising eight surfaces in each corner
octant, which we label $1$--$8$ for later reference, and a central `cross-shaped'
surface.  Far from the origin, the corner surfaces asymptote to the planes $P_i
= \pm 1$.  This is a consequence of the more general property that in the limit
$|P_i|\to\infty$, Eq.~\eqref{eq:3 phase f} reduces to the two-dimensional level
set corresponding to the equivalent two-phase problem with phase $i$ removed.
This also explains the extended stems of the central cross, since slices of the
surface in the far field limits $|P_2|\to \infty$ and $|P_3|\to\infty$
(removing either of the fluid phases) may
be compared with the two-phase level set in figure~\ref{fig:pq
plot}(\emph{a}), with the stems of the cross giving rise to the closed curve
around the origin. 

\begin{figure}
    \includegraphics[width=\textwidth]{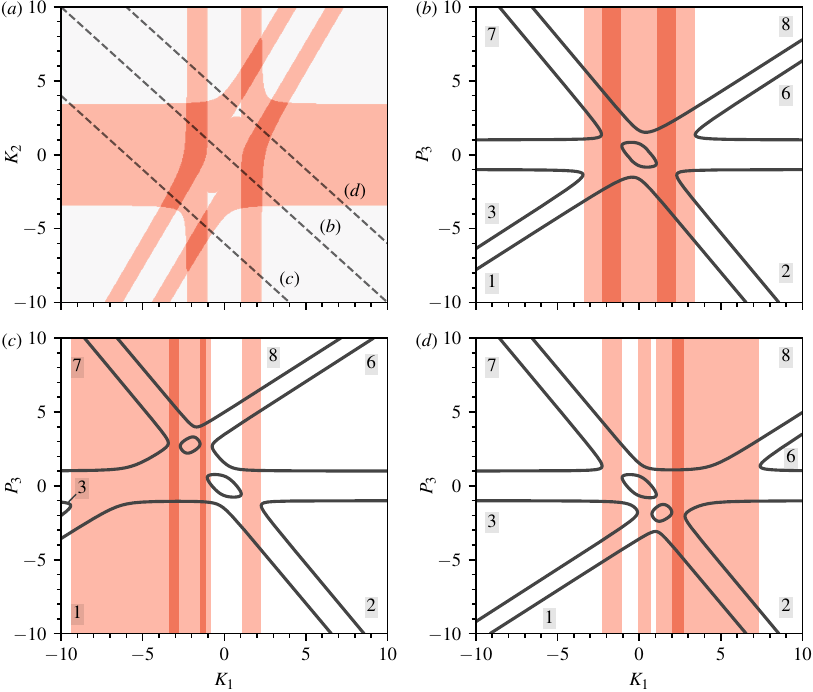}%
    \caption{(\emph{a})~Regions of the $(K_1,K_2)$ plane, for which the
    ($R_{H_1} = R_{H_2}=1$) surface geometry in figure~\ref{fig:pqr} gives rise to
    $6$~(white), $4$~(pink) or $2$~(red) real eigenvalues.
    (\emph{b--d})~Intersections between the Eq.~\eqref{eq:pqr line} line and the
    Eq.~\eqref{eq:3 phase f} level surface (solid lines), for $(K_1, K_2)$
    values along the corresponding dashed lines plotted in panel~(\emph{a}).
    These are: (\emph{b})~$K_2 = -K_1$, (\emph{c})~$K_2 = -K_1-6$ and
    (\emph{d})~$K_2 = -K_1+4$. The shaded bands indicate the number of
    intersections, in accordance with the colouring in~(\emph{a}).
    Labels denote regions enclosed by the numbered corner surfaces (see
    figure~\ref{fig:pqr}).
    }%
    \label{fig:3phase illp}%
\end{figure}

Also plotted in figure~\ref{fig:pqr} is the corresponding line with $K_1 = K_2 =
0$, which passes through $(0,0,0)$.  Using the fact that
$\beta_{21}=\beta_{22}=\beta_{23}$ and $\beta_{31}=\beta_{32}=\beta_{33}$ for
this model, it may be verified that the origin is an additional isolated point
on the level set. Since the line also necessarily passes through the central
cross surface and the corner surfaces $1$ and $8$, it intersects with the level set five times in total. Therefore, this
case corresponds to a repeated real root of $p_c$. More generically, we should
expect an even number of purely real eigenvalues, determined by the number of
intersections of the plotted line through the origin after undertaking an
appropriate translation in the $(P_1,P_2)$ plane by $(K_1, K_2)$, depending on the
values of $\Fr(R_{u_i}-1)$ at a given point in parameter space.
The different possibilities are summarised in figure~\ref{fig:3phase illp}.
In particular, figure~\ref{fig:3phase illp}(\emph{a}) plots, as a function of
$(K_1, K_2)$, 
whether there are $6$~real roots (white), $4$~real roots (pink) or only $2$
(red).  Cases where there
are repeated real roots are either associated with tangential intersections
between the line and level set, [the borders of each shaded region in
figure~\ref{fig:3phase illp}(\emph{a})], or isolated points such as the origin.
Regions with complex eigenvalues cover a substantial part of the plane.
Notably, the model is ill posed as an initial value problem in this case for
all $(K_1,K_2)\in [-1,1]^2$, i.e.\ when
\begin{subequations}
\begin{equation}
    |R_{u_1} - 1| \leq \sqrt{\beta_{11}R_{H_1}}
    \quad \mathrm{and} \quad
    |R_{u_2} - 1| \leq \sqrt{\beta_{22}R_{H_2}}.
    \tag{\theequation\emph{a,b}}%
    \label{eq:Ru1Ru2 small}%
\end{equation}
\end{subequations}
From the geometric picture in figure~\ref{fig:pqr}, we see that this is because
there are only $4$ available intersections in these cases (excepting the special
case $R_{u_1}=R_{u_2}=1$, already discussed).
This observation contrasts with the two-phase models analysed above, which are
always well posed when $R_u$ is sufficiently close to unity.

There are various other possibilities when $R_{u_1}$ or $R_{u_2}$ are large
enough to lie outside the intervals in~\eqref{eq:Ru1Ru2 small}.  The diagrams in
figures~\ref{fig:3phase illp}(\emph{b--d}) are useful for visualising them.  These
plots show how the intersections of the line and level set change as the line is
translated along different trajectories in the $(K_1,K_2)$ plane, represented by
dashed lines in figure~\ref{fig:3phase illp}(\emph{a}).  The first of these, in
figure~\ref{fig:3phase illp}(\emph{b}), considers translations with $K_2 = -K_1$.
When $|K_1| < 1.06$~(3 s.f.), four intersections are identified, as already
discussed.  However, when $1.06 < |K_1| < 2.30$~(3~s.f.), two intersections are
lost, since the line no longer passes through the central cross-shaped surface.
On increasing $|K_1|$ further, two pairs of intersections are created with
corner surfaces (2 and~6 for $K_1 > 2.30$, 3~and~7 for $K_1 < -2.30$),
ultimately leading to well-posed regions when $|K_1| > 3.43$~(3~s.f.).
figure~\ref{fig:3phase illp}(\emph{c}) shows translations with $K_2 = -K_1 -6$.
In this case, when $K_1 = 0$, the line passes out of region~1, through the arm
of the cross that extends along the $P_1 = 0$ plane and clips the 6th corner
surface before passing into region 8, leading to $6$ intersections. Larger $K_1$
values lead to a band of complex characteristics ($1.04 < K_1 < 2.26$,~3~s.f.),
where the line misses the cross arm.  When $K_1$ is lowered from zero, it misses
region 6 and the cross arm in turn, leaving only $2$ intersections for $-1.48 <
K_1 < -1.04$~(3~s.f.).  In the interval $-2.69 < K_1 < -1.48$~(3~s.f.), the
line again
intersects with the cross surface, this time through the arm extending along the
$P_3$-axis.  Lowering $K_1$ further leads to intersections with the $7$th and
$3$rd corner surfaces for $K_1 < -3.42$~(3~s.f.) and $K_1 < -9.40$~(3~s.f.)
respectively.  Finally, the intersections depicted in figure~\ref{fig:3phase
illp}(\emph{d}), which cover translations along $K_2 = -K_1 + 4$, are similar,
but highlight an additional case: for $0.786 < K_1 < 1.05$~(3~s.f), the line
clips through both arms of the cross, leading to a small well-posed band.
Translations farther from the origin can also lead to intersections with
regions $4$ and $5$.

Varying $R_{H_1}$ and $R_{H_2}$ alters both the line and the level set.
However, our earlier observation that the limits $|P_i|\to\infty$ reduce to
two-phase models implies that the resulting parameter space must always contain
ill-posed regions. This is likewise true for any model of the form given by
Eq.~\eqref{eq:3 phase jacobian}.  Therefore, while there may exist other
three-phase models that possess more favourable properties near the origin [a
fully general analysis would require us to classify all surfaces of the form
given in Eq.~\eqref{eq:3 phase f}], none of these systems can be unconditionally
well posed.  Returning to the example case of the~\cite{Pudasaini2019} model, an
investigation of different values in the ranges $R_{H_1},R_{H_2}\in
[0.1,10]$ led to regions of complex characteristics that qualitatively
match the plot in figure~\ref{fig:3phase illp}(\emph{a}), suggesting that the
observations made thus far are robust across parameter space.
It should be noted that in its full generality, this model also includes the
option to include added mass forces and diffusive stresses.  Though the presence
of former terms alters the system's characteristic structure, the
\S\ref{sec:added mass} analysis of the corresponding two-phase case does
little to suggest that they will substantially improve matters.  The effect of
diffusion is dealt with in the next section.

\section{Regularisation}
\label{sec:regularisation}%
The question of how best to alleviate the ill posedness in these models is
fraught with difficulty. Its presence in model equations is usually attributed
to neglected physical effects~\citep{Joseph1990}. For example, in the related
case of two-layer fluid models, the emergence of complex characteristics has
been linked to the impossibility of resolving the vertical mixing induced by
Kelvin-Helmholtz instabilities within a depth-averaged set of governing
equations~\citep{Castro2001}. However, the physics of debris flows are far from settled
and the relative importance of neglected effects may depend on the specifics of
a particular flow. Moreover, it is highly challenging to measure
debris flows in situ, which removes the possibility of examining interior flow
instabilities.

Nevertheless, an obvious candidate to investigate is longitudinal diffusion of
momentum, since it is already included in some models and provides
a clear mechanism for damping instabilities at high wavenumber. For example,
in the typical case where $\matr{A}=\matr{I}$, it is
straightforward to show that a full rank diffusion matrix $\matr{D}$ in
Eq.~\eqref{eq:full eigenproblem} prohibits $\Real(\sigma)$ from blowing up as
$k\to\infty$, provided all its eigenvalues are positive.  However, since there
is no clear reason to include diffusion in the equations for mass conservation,
$\matr{D}$ generally will not be full rank and a deeper analysis is required.

\subsection{A general framework for finding Hadamard instabilities}
\label{sec:general framework}%
We return to the linear stability problem given in Eq.~\eqref{eq:full
eigenproblem}. A general procedure for detecting the presence or absence of
Hadamard instabilities is developed. Since it is cast as an arbitrary matrix
equation, there is no restriction on the dimensionality $N$ of the system, so
our analysis in this subsection is applicable to models with any number of
phases $n=N/2$.  Readers that would rather skip the linear algebra may proceed
to the final paragraph of this subsection, where the method for determining 
posedness is recapitulated.

First, we bring Eq.~\eqref{eq:full eigenproblem} into a simpler form for
analysis.
The matrix $\matr{A}$ must be invertible, in
order for there to be $N$ independent time-evolving fields.
Furthermore, 
we assume that the matrix $\matr{A}^{-1}\matr{D}$ is diagonalisable, since this
covers all the specific cases in this paper. Then, the problem
may be reformulated in terms of
a basis
$\{\hat{\vect{e}}_1,\ldots,\hat{\vect{e}}_N\}$ with respect to which
$\matr{A}^{-1}\matr{D}$ is diagonal.  Therefore, for each matrix $\matr{M} \in
\{\matr{B}, \matr{C}, \matr{D}\}$, we define 
\begin{subequations}
\begin{equation}
\hat{\matr{M}} =
    \matr{P}^{-1}\matr{A}^{-1}\matr{M}\matr{P}\quad\mathrm{and}\quad\hat{\vect{v}} =
    \matr{P}^{-1}\vect{v}, 
    \tag{\theequation\emph{a,b}}%
    \label{eq:basis change}%
\end{equation}
\end{subequations}
for any vector $\vect{v}$, where $\matr{P}$ is a basis change matrix that
diagonalises $\matr{A}^{-1}\matr{D}$. With respect to this transformation,
Eq.~\eqref{eq:full eigenproblem} becomes
\begin{equation}
    \sigma \hat{\vect{r}} + \im k \hat{\matr{B}}\hat{\vect{r}}
    = \hat{\matr{C}}\hat{\vect{r}} - k^2 \hat{\matr{D}}\hat{\vect{r}},
    \label{eq:diagonal eigenproblem}%
\end{equation}
with $\hat{\matr{D}}$ a diagonal matrix.
At high wavenumber $k\gg 1$, we make the following asymptotic expansions:
\begin{subequations}
\begin{equation}
    \sigma = -\sigma_2 k^2 - \im\sigma_1 k + \sigma_0 + \ldots, \quad
    \hat{\vect{r}} = \hat{\vect{r}}_0 + k^{-1}\hat{\vect{r}}_{-1} + \ldots,
    \tag{\theequation\emph{a,b}}%
    \label{eq:asym expansion 1}%
\end{equation}
\end{subequations}
substitute them into Eq.~\eqref{eq:diagonal eigenproblem} and look for the
leading-order terms.  Therefore, at $O(k^2)$, the problem reduces to
\begin{equation}
    \hat{\matr{D}}\hat{\vect{r}}_0 = \sigma_2 \hat{\vect{r}}_0.
    \label{eq:O(k^2)}%
\end{equation}
Noting the sign convention in Eqs.~\eqref{eq:asym expansion 1}, the eigenvalues
$\sigma_2$, which represent diffusion coefficients for the linear problem, must
each have non-negative real part in order to avoid blow-up of $\Real(\sigma)$.
The growth of modes with $\sigma_2 = 0$ is determined beyond this leading order
balance.
If $\hat{\matr{D}}$ is not full rank, it has $i \in\{1,\ldots, N\}$ zero
eigenvalues. Without loss of generality, we locate these in the first $i$
diagonal values of $\hat{\matr{D}}$. The corresponding eigenvectors are
determined only up to an $i$-dimensional subspace ($\hat{\vect{r}}_0 \in
\spn\{\hat{\vect{e}}_1,\ldots,\hat{\vect{e}}_i\}$), by Eq.~\eqref{eq:O(k^2)}.

Therefore, we proceed to the $O(k)$ part of the asymptotic expansion of
Eq.~\eqref{eq:diagonal eigenproblem}. When $\sigma_2 = 0$, this is
\begin{equation}
    (\hat{\matr{B}} - \sigma_1 \matr{I})\hat{\vect{r}}_0
    =
    \im\hat{\matr{D}}\hat{\vect{r}}_{-1}.
    \label{eq:O(k)}
\end{equation}
Since $\hat{\vect{r}}_0 \in \spn\{\hat{\vect{e}}_1,\ldots,\hat{\vect{e}}_i\}$,
only the first $i$ columns of $\hat{\matr{B}}-\sigma_1\matr{I}$ enter into this
system of equations on the left-hand side.  Furthermore, only the first $i$ rows
of Eq.~\eqref{eq:O(k)} are needed to determine $\hat{\vect{r}}_0$ and these are
rows for which the right-hand side is zero.  Consequently, the $\sigma_1$ values
are the eigenvalues of the matrix $\hat{\matr{B}}$ with the last $N-i$ rows and
columns removed.  We shall write $\matr{M}_{\mathrm{red}}$ to denote any matrix
$\matr{M}$ reduced in this way by deleting rows and columns associated with the
nullspace of the diagonal matrix $\hat{\matr{D}}$.  Referring back to
Eqs.~\eqref{eq:asym expansion 1}, we obtain a second criterion that must be met
to avoid Hadamard instability: the eigenvalues $\sigma_1$ of
$\hat{\matr{B}}_{\mathrm{red}}$ must be real.  If these values are also
distinct, then the growth rates stay bounded as $k \to \infty$.

However, $\hat{\matr{B}}_{\mathrm{red}}$ 
may have repeated eigenvalues, which can also lead to blow-up of
$\Real(\sigma)$.  To see why,
we proceed to the $O(1)$ equation with $\sigma_2 = 0$, which reads
\begin{equation}
    (\sigma_0 \matr{I} - \hat{\matr{C}})\hat{\vect{r}}_0
    + \im(\hat{\matr{B}} - \sigma_1 \matr{I})\hat{\vect{r}}_{-1}
    =
    -\hat{\matr{D}}\hat{\vect{r}}_{-2}.
    \label{eq:O(1)}%
\end{equation}
To eliminate dependence of the left-hand side on the unknown
vector $\hat{\vect{r}}_{-2}$, the left eigenvectors,
corresponding to the eigenproblem adjoint to Eq.~\eqref{eq:diagonal
eigenproblem}, may be used. By repeating the arguments used to determine
$\hat{\vect{r}}_0$, 
these may be expanded as $\hat{\vect{l}} = \hat{\vect{l}}_0 +
k^{-1}\hat{\vect{l}}_{-1}+\ldots$
and inferred to satisfy
$\hat{\vect{l}}_0^T\hat{\matr{D}} = \vect{0}$ and
$\hat{\vect{l}}_0^T(\hat{\matr{B}} - \sigma_1 \matr{I}) = \im
\hat{\vect{l}}_{-1}^T\hat{\matr{D}}$ (when $\sigma_2 = 0$).
For any of the $i$ modes, the dot product of the leading order left eigenvector
$\hat{\vect{l}}_0$ may be taken with Eq.~\eqref{eq:O(1)} and
on rearranging the result, the formula
\begin{equation}
    \sigma_0 = \frac{\hat{\vect{l}}_0 \cdot
    \hat{\matr{C}}\hat{\vect{r}}_0
    +\hat{\vect{l}}_{-1}\cdot\hat{\matr{D}}\hat{\vect{r}}_{-1}}{\hat{\vect{l}}_0\cdot\hat{\vect{r}}_0}
    \label{eq:O(1) growth}%
\end{equation}
is obtained. Note that the relevant components of $\hat{\vect{l}}_{-1}$
and $\hat{\vect{r}}_{-1}$ 
required to compute the second term in the numerator
are fully determined by inverting the final $N-i$
rows of~\eqref{eq:O(k)} and their adjoint counterparts. The
left and right eigenvectors for
$\hat{\matr{B}}_{\mathrm{red}}$ are the vectors $\hat{\vect{l}}_0$,
$\hat{\vect{r}}_0$ with the last $N-i$ entries (which are all zeros) deleted.
When $\hat{\matr{B}}_{\mathrm{red}}$ is diagonalisable, these vectors form a
biorthonormal set, with the left and right eigenvectors for each mode satisfying
$\hat{\vect{l}}_0\cdot\hat{\vect{r}}_0 = 1$, so the $O(1)$ growth rate in
Eq.~\eqref{eq:O(1) growth} is well defined.
However, if $\hat{\matr{B}}_{\mathrm{red}}$ is not diagonalisable, at least one
of its eigenvalues is defective.  Therefore, $\sigma_1$ is a repeated eigenvalue
associated with one or more Jordan chains of length at least two.  Then for the
full matrix $\hat{\matr{B}}$ there are two pairs of corresponding generalised
left and right eigenvectors $\hat{\vect{l}}_{0,1}$, $\hat{\vect{l}}_{0,2}$ and
$\hat{\vect{r}}_{0,1}$, $\hat{\vect{r}}_{0,2}$ respectively (in
$\spn\{\hat{\vect{e}}_1,\ldots,\hat{\vect{e}}_i\}$), which satisfy
\begin{subequations}
\begin{gather}
    \begin{cases}
        \hat{\vect{l}}_{0,1}^T(\hat{\matr{B}}-\sigma_1 \matr{I}) =
        \im\hat{\vect{l}}_{-1}^T\hat{\matr{D}},\\
    \hat{\vect{l}}_{0,2}^T(\hat{\matr{B}}-\sigma_1 \matr{I}) =
        \hat{\vect{l}}_{0,1}^T +\hat{\vect{\chi}}^T,
    \end{cases}
    \mathrm{and}
    \quad
    \begin{cases}
        (\hat{\matr{B}}-\sigma_1 \matr{I})\hat{\vect{r}}_{0,1} =
        \im\hat{\matr{D}}\hat{\vect{r}}_{-1},\\
    (\hat{\matr{B}}-\sigma_1 \matr{I})\hat{\vect{r}}_{0,2} =
        \hat{\vect{r}}_{0,1} +\hat{\vect{\Gamma}},
    \end{cases}
    \tag{\theequation\emph{a,b}}%
\end{gather}
\label{eq:generalised eigenvecs}%
\end{subequations}
where $\hat{\vect{r}}_{0,1} \equiv \hat{\vect{r}}_0$ and
$\hat{\vect{l}}_{0,1}\equiv \hat{\vect{l}}_0$, and
$\hat{\vect{\chi}},\hat{\vect{\Gamma}}$ are unknown vectors in
$\spn\{\hat{\vect{e}}_{i+1},\ldots,\hat{\vect{e}}_N\}$.  In this case, the
formula in
Eq.~\eqref{eq:O(1) growth} is always singular, since projecting any left
eigenvector onto~(\ref{eq:generalised eigenvecs}\emph{b}) shows that
$\hat{\vect{l}}_0 \cdot \hat{\vect{r}}_0 = 0$.  Physically, this singularity can
be thought to emerge from a resonance between two or more modes that collapse
onto one another when $\hat{\matr{B}}_{\mathrm{red}}$ becomes defective.
Examples of this are given below, in \S\ref{sec:existing models}.

The failure of Eq.~\eqref{eq:O(1) growth} in these cases suggests the need for
an alternative asymptotic expansion.
Anticipating growth of some intermediate order
between $O(k)$ and $O(1)$, we replace the expansions in Eqs.~\eqref{eq:asym expansion 1} 
with
\begin{subequations}
\begin{equation}
    \sigma = -\im\sigma_1 k + \sigma_{1/2} k^{1/2} + \sigma_0 + \ldots, \quad
    \hat{\vect{r}} = \hat{\vect{r}}_{0,1} + k^{-1/2}\hat{\vect{r}}_{-1/2} +
    k^{-1}\hat{\vect{r}}_{-1}+\ldots.
    \tag{\theequation\emph{a,b}}%
\end{equation}
\end{subequations}
This leaves the analysis at $O(k)$ unchanged
and introduces the following equation at $O(k^{1/2})$:
\begin{equation}
    \sigma_{1/2} \hat{\vect{r}}_{0,1} + \im (\hat{\matr{B}} - \sigma_1 \matr{I})
    \hat{\vect{r}}_{-1/2} + \hat{\matr{D}} \hat{\vect{r}}_{-3/2} = \vect{0}.
\end{equation}
We project this onto $\hat{\vect{l}}_{0,2}$ and use Eq.~\eqref{eq:generalised
eigenvecs}, along with the fact that $\hat{\vect{l}}_{0,2}$ is orthogonal to the
range of $\hat{\matr{D}}$, to conclude that
\begin{equation}
    \sigma_{1/2} \hat{\vect{l}}_{0,2}\cdot\hat{\vect{r}}_{0,1} +
        \im(\hat{\vect{l}}_{0,1} +\hat{\vect{\chi}})\cdot
    \hat{\vect{r}}_{-1/2} = 0.
\label{eq:O(sqrt(k)) simplified}
\end{equation}
The unknown vector $\hat{\vect{r}}_{-1/2}$ is eliminated by proceeding to the
$O(1)$ equation. With the new expansion, this is
\begin{equation}
    \sigma_{1/2}\hat{\vect{r}}_{-1/2} + \sigma_0 \hat{\vect{r}}_{0,1}
    +\im(\hat{\matr{B}} - \sigma_1 \matr{I})\hat{\vect{r}}_{-1} 
    - \hat{\matr{C}}\hat{\vect{r}}_{0,1} + \hat{\matr{D}}\hat{\vect{r}}_{-2}
    = \vect{0}.
\end{equation}
Then, we project this onto $\hat{\vect{l}}_{0,1}$. 
Since $\hat{\vect{l}}_{0,1}\cdot\hat{\vect{r}}_{0,1} = 0$, the term containing
$\sigma_0$ vanishes, along with the diffusive term which lies in an orthogonal
subspace.
After rearranging and using
Eq.~\eqref{eq:O(sqrt(k)) simplified}, 
as well as the $O(k^{3/2})$ part of the system, which implies that
$\hat{\vect{r}}_{-1/2}\in\spn\{\hat{\vect{e}}_1,\ldots,\hat{\vect{e}}_i\}$,
we obtain a formula for the $O(k^{1/2})$
part of the growth rate:
\begin{equation}
    \sigma_{1/2} = \pm \frac{1 - \im}{2} \left(
    \frac{2(
    \hat{\vect{l}}_{0,1}\cdot\hat{\matr{C}}\hat{\vect{r}}_{0,1}
    +\hat{\vect{l}}_{-1}\cdot\hat{\matr{D}}\hat{\vect{r}}_{-1})}
    {\hat{\vect{l}}_{0,2}
    \cdot \hat{\vect{r}}_{0,1}
    }
    \right)^{\!1/2}\!\!\!.
    \label{eq:sigma half}%
\end{equation}
For Jordan chains of length two
$\hat{\vect{l}}_{0,2}\cdot\hat{\vect{r}}_{0,1}=|\hat{\vect{l}}_{0,2}||\hat{\vect{r}}_{0,1}|\neq
0$, provided both the left and right vectors correspond to the same Jordan
block.  Consequently, Eq.~\eqref{eq:sigma half} implies that there is a mode
such that $\Real(\sigma) \sim k^{1/2}$, provided the terms in the numerator do
not interact in a way that causes it to vanish. 

Conversely, for longer Jordan chains, the denominator in the Eq.~\eqref{eq:sigma
half} formula is also guaranteed to be singular. Different asymptotic expansions
are needed, depending on the length of the the chain.  However, to avoid these
further complications, we terminate our analysis here, since cases where
three or more modes intersect at high wavenumber are far less commonly
encountered.

To summarise the analysis above, models up to second order that may be cast in
the general form of Eq.~\eqref{eq:gen model eqs} are ill posed as initial-value
problems if any of the following conditions are met:
\begin{enumerate}
    \item Any eigenvalue of $\hat{\matr{D}}$ is negative, where
        $\hat{\matr{D}}$ denotes a diagonalisation of
        $\matr{A}^{-1}\matr{D}$.
    \item Any eigenvalue of $\hat{\matr{B}}_{\mathrm{red}}$ is complex, where
        $\hat{\matr{B}}_{\mathrm{red}}$ denotes the matrix
        formed by representing
        $\matr{A}^{-1}\matr{B}$ in the basis used to diagonalise
        $\matr{A}^{-1}\matr{D}$ in~(i) and deleting each row and column $j$ such
        that the $j$-th diagonal entry of $\hat{\matr{D}}$ is nonzero. We
        refer to $\hat{\matr{B}}_{\mathrm{red}}$ as a `reduced Jacobian' in
        later analysis.
    \item Repeated real eigenvalues of $\hat{\matr{B}}_{\mathrm{red}}$ 
        of algebraic multiplicity $2$ share the same left and right eigenvectors
        $\hat{\vect{l}}_{0,1}$  and $\hat{\vect{r}}_{0,1}$ (up to
        normalisation), and
        the numerator of Eq.~\eqref{eq:sigma half} is nonvanishing.
        [More generally, the expectation following
        from Eq.~\eqref{eq:O(1) growth}, is that repeated real eigenvalues of
        any algebraic multiplicity $m\geq 2$ imply ill posedness if the dimension of their
        associated eigenspace is strictly less than $m$,
        but this is not explicitly proven above.]
\end{enumerate}

For the remainder of this section, we apply these steps to different example
systems.

\subsection{Velocity diffusion in every momentum equation}
\label{sec:full diff well posed}%
Before analysing individual models, we highlight a generic case, which is
guaranteed to be well posed.  Consider an $n$-phase model, with each phase $i$
characterised by height $H_i$ and velocity $\average{u_i}$, organised into pairs
of mass and momentum equations of the form
\begin{subequations}
\begin{equation}
    \frac{\partial H_i}{\partial t} + \frac{\partial~}{\partial x}
    \left(
    H_i \average{u_i}
    \right) = 0, \quad
    \frac{\partial \widebar{u_i}}{\partial t} +
    F_i(H_1,\average{u_1},\ldots,H_n,\average{u_n}) = 0,
    \tag{\theequation\emph{a,b}}%
\end{equation}
\end{subequations}
where the functions $F_i$ contain no dependence on time derivatives or spatial
derivatives of first order or higher.
To each of the
the $j = 1,\ldots,n$ momentum equations, add a term of the form $\frac{\partial~}{\partial
x}(\nu_j(\vect{q})\frac{\partial \widebar{u_j}}{\partial x})$, where
$\nu_j(\vect{q})$ denotes a general diffusivity coefficient function that stays
strictly positive.  
When casting the linearised problem in matrix form, the equations are
ordered so that the mass and momentum equations respectively lie on odd and even
rows, as
before.
The corresponding diffusion matrix is already diagonal, so at any
$\vect{q}=\vect{q}_0$,
\begin{equation}
    \hat{\matr{D}} = \begin{pmatrix}
    0 & 0 & \ldots & 0 & 0  \\
    0 & \nu_1(\vect{q}_0) & \ldots & 0 & 0 \\
    \vdots & \vdots & \ddots & \vdots & \vdots \\
    0 & 0 & \ldots & 0 & 0 \\
    0 & 0 & \ldots & 0 & \nu_n(\vect{q}_0)
\end{pmatrix},
\end{equation}
which is positive semidefinite and damps out growth at high wavenumber for $n$ of
the $2n$ stability modes.  The reduced Jacobian matrix is simply
\begin{equation}
    \hat{\matr{B}}_{\mathrm{red}} =
    \begin{pmatrix}
        \average{u_1} & 0 & \ldots & 0 \\ 
        0 & \average{u_2} & \ldots & 0 \\
        \vdots & \vdots & \ddots & \vdots \\
        0 & 0 & \ldots & \average{u_n} \\
    \end{pmatrix},
\end{equation}
which clearly possesses $n$ real eigenvalues and $n$ linearly independent
eigenvectors. Therefore, debris flow models with $\matr{A}=\matr{I}$ can always
be regularised by adding positive diffusion to every momentum equation.

\subsection{Existing models}
\label{sec:existing models}%
\subsubsection{\cite{Meng2022}}
As detailed in~\S\ref{sec:mengmodel}, in the model of~\cite{Meng2022},
$\matr{A}=\matr{I}$ and only diffusion in the fluid phase is included.  The
equations are~\eqref{eq:pitman le 1}, \eqref{eq:meng 1}, \eqref{eq:pitman
le 3} and~\eqref{eq:meng 2}, rendered dimensionless as described in
\S\ref{sec:two-phase}. The diffusion matrix is already diagonal and is given
by
\begin{equation}
    \hat{\matr{D}} = \begin{pmatrix}
0 & 0 & 0 & 0 \\
0 & 0 & 0 & 0 \\
0 & 0 & 0 & 0 \\
0 & 0 & 0 & 2\nu_f
\end{pmatrix}
,
\end{equation}
where $\nu_f \equiv \eta_f / (\rho_f H_f^{(0)}\widebar{u_f}^{(0)})>0$ is a
dimensionless kinematic viscosity coefficient (though it could equally be viewed
as an eddy diffusivity if the flow is turbulent).
Therefore, the corresponding reduced Jacobian is formed by
removing the fourth row and column from the full matrix $\hat{\matr{B}} =
\matr{B}$, given in Eq.~\eqref{eq:jacobian}. At $\vect{q}=\vect{q}_0$,
\begin{equation}
    \hat{\matr{B}}_{\mathrm{red}} = 
    \begin{pmatrix}
        R_u & R_H & 0 \\
        (\gamma+\beta_1)\Fr^{-2} & R_u & (\gamma + \beta_2)\Fr^{-2} \\
        0 & 0 & 1
    \end{pmatrix}.
    \label{eq:Bred meng}%
\end{equation}
Its eigenvalues are 
\begin{equation}
    \sigma_1 = 1, R_u \pm \frac{\sqrt{R_H(\gamma + \beta_1)}}{\Fr}
\end{equation}
with corresponding eigenvectors
\begin{equation}
\begin{pmatrix}
    R_H \\
    1 - R_u \\
    \frac{(R_u-1)^2{\mathit{Fr}}^2-R_H(\gamma+\beta_1)}{\gamma+\beta_2}
\end{pmatrix},
~~
\begin{pmatrix}
    R_H \\
    \pm \frac{1}{\mathit{Fr}}\sqrt{R_H(\gamma+\beta_1)} \\
    0
\end{pmatrix}.
    \label{eq:evecs}%
\end{equation}
Firstly, note that the latter pair of $\sigma_1$ values equal the
characteristics for the solids phase of the `decoupled' problem, given in
Eqs.~\eqref{eq:uncoupled characteristics}. Hence, all values are expected to be
real. 
However, there is the opportunity for repeated eigenvalues, which
occurs when $\sigma_1=1$ matches either of the other two growth
rates, i.e.\ when
\begin{equation}
    R_u = 1 \pm \frac{1}{\Fr}\sqrt{R_H(\gamma + \beta_1)}.
    \label{eq:meng ill posed line}%
\end{equation}
This is similar, but not equivalent to the condition for intersecting
decoupled characteristics,
given previously in Eq.~\eqref{eq:uncoupled
intersection}.
By substituting Eq.~\eqref{eq:meng ill posed line} into Eq.~\eqref{eq:evecs}, it may be verified that the
corresponding eigenvectors are equal when this condition is satisfied.
Consequently, the equations feature an instability with order $k^{1/2}$ growth
rate in the high-wavenumber asymptotic limit and are
ill posed wherever Eq.~\eqref{eq:meng ill posed line} is satisfied.

Diffusion of momentum is often also included in shallow models of dry granular
flows~\citep[e.g.][]{Gray2014,Baker2016}. The general argument in
\S\ref{sec:full diff well posed} implies that adding a diffusive term of the
form $\frac{\partial~}{\partial x}(\nu_s \frac{\partial \average{u_s}}{\partial
x})$, where $\nu_s \equiv \nu_s(\vect{q}_0) > 0$, to the solids momentum
equation is sufficient to regularise this model.  Moreover, using analogous
arguments to those above, it may be verified that diffusive terms in both
momentum terms are required, in order to guarantee that the model stays
unconditionally well posed.

Specifically, if $\nu_f = 0$, but diffusion in the solids momentum equation is
included, then the reduced Jacobian is formed by eliminating the second row and
column of $\hat{\matr{B}}$, to leave
\begin{equation}
    \hat{\matr{B}}_{\mathrm{red}}=
    \begin{pmatrix}
        R_u & 0 & 0 \\
        0 & 1 & 1 \\
        \Fr^{-2} & \Fr^{-2} & 1
    \end{pmatrix}.
\end{equation}
This matrix is defective when
\begin{equation}
R_u = 1 \pm 1/\Fr,
\end{equation}
giving rise to a family of $O(k^{1/2})$ instabilities at these points in
parameter space, similar to the case where only fluid diffusion is included.

Note that since we used the general form of $\matr{B}$ from
Eq.~\eqref{eq:jacobian} to construct the reduced Jacobians, these assessments
apply also to the case of adding simple diffusive terms to regularise the models
of~\cite{Pitman2005},  \cite{Pelanti2008} and~\cite{Meyrat2022}. 

\subsubsection{\cite{Pudasaini2012}}
\label{sec:pudasaini diffusive}%
This model incorporates two diffusive stresses for the fluids phase: a
Newtonian component, equivalent to the term used by~\cite{Meng2022} and a
non-Newtonian closure defined in~\S\ref{sec:pudasaini}.
The relevant contributions to the depth-averaged downslope momentum equation are
the second-order terms of Eq.~\eqref{eq:pudasaini fluid stress av}.  When the
model is converted to the quasilinear form that was used for the local analysis,
the fluid momentum equation is non-dimensionalised (as
per~\S\ref{sec:two-phase}) and divided through by $\rho_f H_f$, and the
diffusive terms become
\begin{equation}
    \frac{2\nu_f}{H_f}\frac{\partial ~}{\partial x}\left(
    H_f\frac{\partial \average{u_f}}{\partial x}
    \right)
    +
    \frac{2 \nu_f \mathcal{N}}{H_f}
    \frac{\partial~}{\partial x}
    \left[
        H_f(\average{u_s} - \average{u_f})
    \frac{\partial~}{\partial x}\left(
    \frac{H_s}{H_s + H_f}
    \right)
    \right]\!.
    \label{eq:pudasaini diffusion}%
\end{equation}
The parameter $\mathcal{N} = \average{{\mathcal{A}}}/\average{\varphi_f}$ is a
ratio of the effective diffusion coefficients for the Newtonian and
non-Newtonian parts [see Eq.~\eqref{eq:pudasaini fluid stress
av}] and is assumed to be constant by~\cite{Pudasaini2012}.

After linearising around $\vect{q}=\vect{q}_0 = (R_H, R_u, 1, 1)^T$\!, 
the diffusion matrix becomes
\begin{equation}
    \matr{D}=
    2\nu_f
    \begin{pmatrix}
        0 & 0 & 0 & 0 \\
        0 & 0 & 0 & 0 \\
        0 & 0 & 0 & 0 \\
        \frac{\mathcal{N}(R_u-1)}{(1+R_H)^2} & 0 &
        \frac{\mathcal{N}R_H(1-R_u)}{(1+R_H)^2} & 1
    \end{pmatrix}
    .
    \label{eq:pudasaini D}%
\end{equation}
The matrices $\matr{A}$ and $\matr{B}$ were given previously, in
Eqs.~\eqref{eq:pudasaini A} and~\eqref{eq:pudasaini B} respectively.
The basis change matrix 
\begin{equation}
    \matr{P}=
    \begin{pmatrix}
        1 & 0 & 0 & 0 \\
        0 & 1 & 0 & \gamma\average{C} \\
        0 & 0 & 1 & 0 \\
        \frac{\mathcal{N}(1-R_u)}{(1+R_H)^2} & 0 &
        \frac{\mathcal{N}R_H(R_u-1)}{(1+R_H)^2} & \gamma\average{C}+1
    \end{pmatrix}
    ,
    \label{eq:pudasaini P}%
\end{equation}
diagonalises $\matr{A}^{-1}\matr{D}$, so that
$\hat{\matr{D}}=\matr{P}^{-1}\matr{A}^{-1}\matr{D}\matr{P}$ is the matrix of all
zeros, save for its only eigenvalue $\sigma_2$, located on the bottom right
entry,
\begin{equation}
    \mathsfi{D}_{44} = \sigma_2 = \frac{2\nu_f(\gamma\average{C} + 1)}{1 +
    \average{C}(\gamma+R_H)}.
\end{equation}
Since $\sigma_2 > 0$, there is no blow-up at $O(k^2)$ and it remains to check
the properties of $\hat{\matr{B}}_{\mathrm{red}}$, which is formed by removing
the fourth row and column of $\matr{P}^{-1}\matr{A}^{-1}\matr{B}\matr{P}$.

In the general case, analytical expressions for this matrix are cumbersome and
it is better to compute its eigenvalues numerically. However, two limiting cases
are tractable.  Firstly, when the non-Newtonian viscosity is not included,
$\mathcal{N}=0$ and the reduced Jacobian is
\begin{equation}
    \hat{\matr{B}}_{\mathrm{red}} =
\begin{pmatrix}
    R_u & R_H & 0 \\
    \frac{1}{\gamma\average{C}+1}\left[\frac{\gamma + \beta_1}{{\small{\Fr}}^{2}}
    +\frac{\average{C}(R_u-1)}{R_H}\right]& R_u +
    \frac{\gamma\average{C}}{\gamma\average{C}+1} & \frac{\gamma +
    \beta_2}{(\gamma\average{C}+1){\small{\Fr}}^2} \\
    0 & 0 & 1
\end{pmatrix}.
\end{equation}
The eigenvalues $\sigma_1$ of this matrix are
\begin{equation}
    \sigma_1 = 1, 
    R_u + \frac{1}{\gamma\average{C}+1}\left(
    \frac{\gamma\average{C}}{2} \pm \frac{\sqrt{\Delta}}{\Fr}
    \right),
    \label{eq:A=0 sigma1}%
\end{equation}
where
\begin{equation}
    \Delta = 
    (\gamma\average{C}+1)
    \left[
        \Fr^2\gamma\average{C}(R_u-1)
        +(\gamma+\beta_1)R_H
        \right]
    +\left(\frac{\Fr\hspace{0.1em}\gamma\average{C}}{2}\right)^2\!\!.
\end{equation}
The latter pair are complex conjugate iff $\Delta < 0$. Rearranging this
inequality leads to
\begin{equation}
    R_u - 1 < -\frac{\gamma\average{C}}{4(\gamma\average{C}+1)}
    - \frac{R_H(\gamma+\beta_1)}{\gamma\average{C}\Fr^2}.
    \label{eq:A=0 inequality}%
\end{equation}
This describes a region of complex eigenvalues that is constrained to lie within 
$R_u < 1$.
Note that in the $\average{C}\to 0$ limit, this region entirely recedes and
inequality~\eqref{eq:A=0 inequality} is never satisfied.
In addition to these complex eigenvalues, there is the opportunity for
$O(k^{1/2})$ blow-up if $\hat{\matr{B}}_{\mathrm{red}}$ is defective, which can
happen if $\sigma_1=1$ intersects with either of the other two eigenvalues in
Eq.~\eqref{eq:A=0 sigma1}. The condition for this simplifies to
\begin{equation}
    R_u = 1 \pm
    \frac{1}{\Fr}\sqrt{\frac{R_H(\gamma+\beta_1)}{\gamma\average{C}+1}},
    \label{eq:pudasaini O(k^1/2) line}%
\end{equation}
which generalises Eq.~\eqref{eq:meng ill posed line} for cases where
$\average{C}\geq 0$.  It may be separately verified that only one eigenvector
corresponding to $\sigma_1=1$ exists when $R_u$ satisfies
Eq.~\eqref{eq:pudasaini O(k^1/2) line}, implying that
$\hat{\matr{B}}_{\mathrm{red}}$ is defective here.

\begin{figure}
    \includegraphics[width=\textwidth]{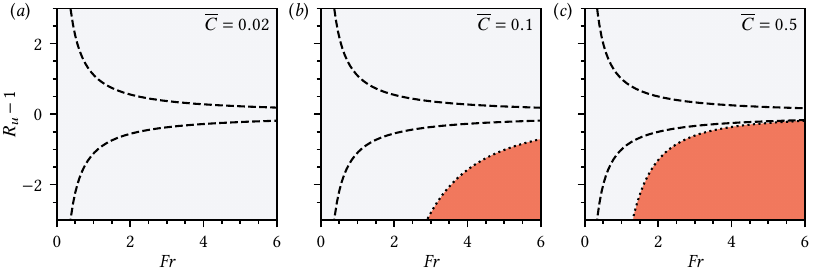}%
    \caption{Regions where the reduced Jacobian for the model
    of~\cite{Pudasaini2012} with diffusive terms possesses complex eigenvalues 
    (red shading), for
    $R_H = 1$, $\gamma = 0.5$,
    $\mathcal{N}=0$ and $\average{C} =$~(\emph{a})~$0.02$, (\emph{b})~$0.1$,
    (\emph{c})~$0.5$.  
    The boundaries of these regions are given analytically by inequality~\eqref{eq:A=0
    inequality} (dotted black).
    Along the black dashed lines, given by Eq.~\eqref{eq:pudasaini O(k^1/2) line},
    the reduced Jacobian is defective.
    The model is ill posed as an initial value problem for flow states that pass
    through either the dashed line, or the red region.
    }%
    \label{fig:pudasaini C effect}%
\end{figure}
In figures~\ref{fig:pudasaini C effect}(\emph{a--c}), we show the regions where
the model is ill posed for $\mathcal{N}=0$, $\average{C}=$ (\emph{a})~$0.02$,
(\emph{b})~$0.1$, (\emph{c})~$0.5$ and
the same
illustrative parameters used in figure~\ref{fig:pudasaini posedness}(\emph{a}).
Dashed curves show the lines given by Eq.~\eqref{eq:pudasaini O(k^1/2)
line}.
The ill-posed region that emerges at low $R_u$ values via inequality~\eqref{eq:A=0 inequality},
whose border is given by the dotted line, may be compared to
similar regions present
in the problem without diffusion, plotted in
figure~\ref{fig:pudasaini posedness}.

\begin{figure}
    \includegraphics[width=\textwidth]{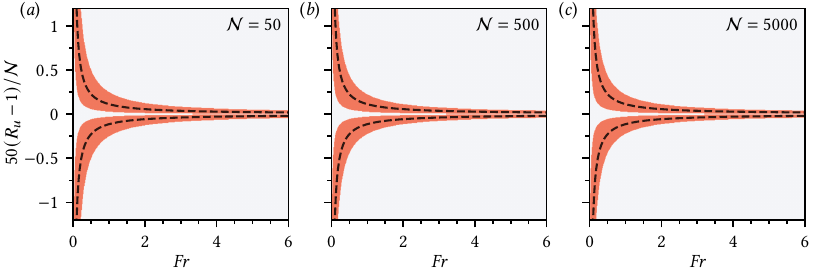}%
    \caption{Regions where the model of~\cite{Pudasaini2012} with diffusive
    terms is ill posed as an initial value problem (red shading), for $R_H = 1$,
    $\gamma = 0.5$, $\average{C}=0$ and high values of the ratio $\mathcal{N}$
    between non-Newtonian and Newtonian diffusion coefficients, $\mathcal{N}
    =$~(\emph{a})~$50$, (\emph{b})~$500$ and (\emph{c})~$5000$.  
    Note that each vertical axis is scaled with respect to $50/\mathcal{N}$ and
    that the shaded regions are near identical under this rescaling.
    Asymptotic
    expansions for the eigenvalues at high $\mathcal{N}$ intersect along the
    black dashed lines, whose formulae are given in Eq.~\eqref{eq:high N
    lines}.}
    \label{fig:pudasaini A effect}%
\end{figure}

If instead, $\mathcal{N}>0$ and the limit of vanishing added mass
$\average{C}\to 0$ is taken, then
\begin{equation}
    \hat{\matr{B}}_{\mathrm{red}} =
\begin{pmatrix}
    R_u & R_H & 0 \\
    (\gamma + \beta_1)\Fr^{-2} & R_u & (\gamma + \beta_2)\Fr^{-2} \\
    \frac{\mathcal{N}(1-R_u)}{(1+R_H)^2} & 0 & 1 +
    \frac{\mathcal{N}R_H(R_u-1)}{(1+R_H)^2}
\end{pmatrix}.
\end{equation}
Note that this is a generalisation of the reduced Jacobian in Eq.~\eqref{eq:Bred
meng}. When the non-Newtonian terms are included,
$\mathcal{N}$ is expected to be a large number compared with the other
parameters [\cite{Pudasaini2012} uses $\mathcal{N} = 5000$].  By solving for
roots of the characteristic polynomial via series expansion, when
$\mathcal{N}\gg 1$, the eigenvalues of $\hat{\matr{B}}_{\mathrm{red}}$ may be
obtained:
\begin{equation}
    \sigma_1 = \frac{(R_u-1)R_H}{(1+R_H)^2}\mathcal{N} + 1 +
    O(\mathcal{N}^{-1}),
    ~~
    \pm \frac{1}{\Fr}\sqrt{(\gamma+\beta_1)R_H + \gamma + \beta_2}
    + O(\mathcal{N}^{-1})
\end{equation}
These expressions are real-valued and remain so in the limit.
However, either of the second and third branches
merges with the first when
\begin{equation}
    R_u = 
    1 \pm \frac{(1+R_H)^2}{\Fr R_H\mathcal{N}}\sqrt{(\gamma+\beta_1)R_H +
    \gamma + \beta_2} + O(\mathcal{N}^{-2}).
    \label{eq:high N lines}%
\end{equation}
As we have seen previously, the merging of branches can give rise to complex
eigenvalues. In this case, the merger originates in the limit
$\mathcal{N}\to\infty$.
For large but finite $\mathcal{N}$, we
compute the eigenvalues of $\hat{\matr{B}}_{\mathrm{red}}$ numerically and
summarise their type in figure~\ref{fig:pudasaini A effect}.  
The two parametric lines given in Eq.~\eqref{eq:high N lines} are flanked by
bands where $\sigma_1$ takes complex values. Furthermore, the shape of these
bands is self-similar in the asymptotic high $\mathcal{N}$ regime.

\subsubsection{\cite{Pudasaini2019}}
\label{sec:three phase diffusive}%
To conclude this section, we touch upon the three-phase model
of~\cite{Pudasaini2019}, which was introduced in~\S\ref{sec:three phase}. It
was shown previously that omitting the
diffusive terms in this model can lead to ill-posed initial value problems.  
However, it
remains to be seen whether including the terms can eliminate this issue.  As
before, we neglect the complications of the added mass effect, though as we have
just seen, this can be analysed using the same methods.

Diffusion of momentum in the~\cite{Pudasaini2019} model appears in the equations
for both fluid phases (which are labelled $2$ and $3$ in \S\ref{sec:three
phase}). Each contains a Newtonian and non-Newtonian component similar to
the terms in~\eqref{eq:pudasaini diffusion} for the~\cite{Pudasaini2012} system. 
The diffusion matrix $\matr{D}$ for the non-dimensionalised and linearised model
equations is given explicitly in Appendix~\ref{appendix:3 phase}. Due to the non-Newtonian terms, it has
off-diagonal entries.
It possesses two non-zero eigenvalues, which are: $2\nu_2 > 0$ and
$2\nu_3>0$, where $\nu_2$, $\nu_3$ are the Newtonian diffusion coefficients associated
with the second and third phases. A suitable basis change matrix $\matr{P}$ that
diagonalises $\matr{D}$ was determined using computational algebra and is
also specified in Appendix~\ref{appendix:3 phase}.
This allows us to form the reduced Jacobian (a $4\times 4$ matrix in this case)
numerically and compute its eigenvalues.

Since there are five independent dimensionless variables $(R_{H_1}, R_{H_2},
R_{u_1}, R_{u_2}, \Fr)$ that specify a particular state (in addition to several
fixed model parameters), we do not attempt an exhaustive study.  Instead, we fix
$R_{H_1}=1$ and investigate the effect of introducing the intermediate fluid
phase by increasing $R_{H_2}$ from zero.  Guided by our analysis in
\S\ref{sec:three phase}, we shift $R_{u_2}$ slightly away from unity, setting
$R_{u_2}=1.01$, to allow for richer interactions between the phases.
Figure~\ref{fig:pm computations} shows the results of these computations, using
illustrative model parameters, given in Appendix~\ref{appendix:3 phase}. These
parameters were selected to match our choices for computations relating to
the~\cite{Pudasaini2012} model with $\mathcal{N}=5000$ [\S\ref{sec:pudasaini
diffusive}], so that when $R_{H_2}\to 0$, the system collapses to the this
two-phase case.
\begin{figure}
    \includegraphics[width=\textwidth]{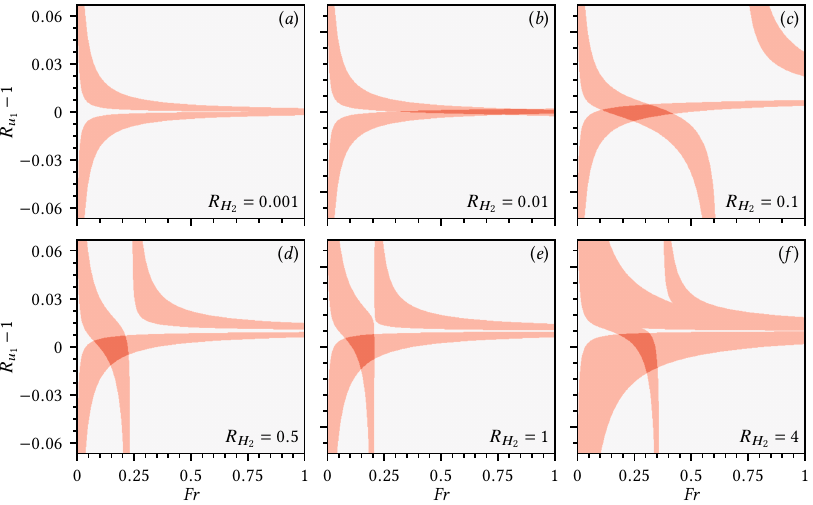}%
    \caption{Illustrative computations of the eigenvalues of 
    $\hat{\matr{B}}_{\mathrm{red}}$ for the model of~\cite{Pudasaini2019},
    with $R_{H_1} = 1$ and $R_{u_2} = 1.01$.  In regions shaded pink, the model
    possesses a single pair of complex eigenvalues, while red shading covers
    areas where two complex pairs were found. Elsewhere, all eigenvalues are
    real. The parameters for these
    computations are given in Appendix~\ref{appendix:3 phase}.
    }%
    \label{fig:pm computations}%
\end{figure}
When $R_{H_2} = 10^{-3}$ [panel~(\emph{a})], there are two bands where the
reduced Jacobian features a pair of complex eigenvalues, either
side of $R_{u_1}=1$. As expected, these closely match the corresponding regions
plotted in figure~\ref{fig:pudasaini A effect}(\emph{c}).  However, the reflection
symmetry of these bands about $R_{u_1} = 1$ is broken for any $R_{H_{2}}>0$.
This becomes apparent at higher $R_{H_2}$ values.  Figures~\ref{fig:pm
computations}(\emph{b,c}) show the cases $R_{H_2} = 0.01$ and $ 0.1$
respectively. The upper band drops below $R_{u_1} < 1$ and overlaps
the lower band, with the Froude numbers at which this occurs decreasing
as $R_{H_2}$ increases. Where the bands overlap, there are two pairs of complex
eigenvalues.
Additionally, a second upper band appears at higher $\Fr$ and draws toward lower
$\Fr$ as $R_{H_2}$ increases to $0.5$ and $1$, in figures~\ref{fig:pm
computations}(\emph{d}) and~(\emph{e}) respectively.
In the final plot [figure~\ref{fig:pm computations}(\emph{f})], at $R_{H_2} = 4$,
the two upper bands have merged, though in this case the merger does not double
the number of complex eigenvalues present.

Other choices for the flow variables lead to plots that are
similar to figure~\ref{fig:pm computations}, at least in the sense that they are
constructed from complicated tangles of complex eigenvalue regions.
While it may be possible to make sense of these diagrams in detail, this
is perhaps beside the point. It is clear, even from this cursory
investigation that this three-phase model suffers from the same issues as the
two-phase models.

\section{Discussion}
We have seen that depth-averaged debris flow models with mass and momentum
equations for more than one phase lead to initial value problems that are only
conditionally well posed.  In particular, they are overcome by catastrophic
instabilities if their flow fields stray into certain regions of parameter
space.  This limits their applicability to cases where solutions provably avoid
these regions. For example, travelling wave solutions, such as those constructed
by~\cite{Meng2022} in their model, are mathematically constrained to have equal
depth-averaged velocities for both phases ($R_u = 1$) -- a case which we have
seen is guaranteed to be well posed for the simplest two-phase models.  However,
since the ill-posed regions lie well within physically accessible regimes, no
such guarantee can be made \emph{a priori} for simulations of real flows over complex
topographies.  This calls into question the reliability of computational results
obtained with multi-phase models in prior studies over the past two decades.
Furthermore, it strongly suggests that these systems should not be used in
scientific applications such as hazards assessment, since any numerical
`solutions' whose flow fields stray into an ill-posed region become impossible
to converge to values that faithfully approximate the underlying partial
differential equations. 

A common observation in our analysis has been that adding physical detail to a
debris flow model can exacerbate the problem of ill posedness by increasing the
opportunities for unwelcome resonant interactions between flow fields.
Therefore, for operational purposes, it may be wisest for practitioners to
adopt a philosophy of favouring models that are `as simple as possible (but no
simpler)'.
In most cases, this will mean
depth-averaged systems
that provide a single bulk momentum equation for the flowing mixture. 
Such systems can capture most of the important debris-flow physics and available
models either inherit well posedness from the classical shallow water equations,
or have independently been shown to be strictly
hyperbolic, such as the models of~\cite{Kowalski2013} and~\cite{George2014}.  
However, simplicity comes with the potential risk of missing or
mispredicting key phenomena,
such as the longitudinal separation of the phases over the length of a debris
flow.
For situations where a fully
multi-phase description is absolutely necessary, careful model development is
needed to resolve the issues raised herein.  

The analysis of \S\ref{sec:general
framework}, which provides a general procedure for identifying ill posedness in
initial value problems of up to second order in their spatial derivatives,
should prove useful in this regard. This may be applied either numerically or
(ideally) analytically, to assess particular models and indicate possible ways
to regularise them.  One option that we have highlighted is to add diffusive
terms.
It is surely reasonable to justify the presence of momentum diffusion in any phase of
a debris flow, over a suitable range of scales and doing so provides a
potentially straightforward way to avoid model pathologies. 
However, while diffusion might be expected to automatically regularise the
system, we show in \S\ref{sec:existing models} that this is not the case for the
existing models analysed herein.
Moreover, the appropriate size of the
diffusion coefficients in each case may not be clear in advance
and careful work is needed in order to formulate these terms rigorously for
particular flows.
Nevertheless, ill posedness is provably avoided 
for the natural case
of a diagonal diffusion operator
with strictly positive
entries for each momentum equation and zeros elsewhere.
Alternatively, it may be possible to benefit from the existing research on
numerical methods for the multi-layer shallow water
equations~\citep{CastroDiaz2023} to design schemes
that avoid non-hyperbolic regimes without diffusive terms,
bearing in mind that any such approach would need to be
physically justified for debris flows.

A deeper question remains. To what extent does the presence of ill posedness in
these models signify the existence of underlying physical instabilities?  The
removal of the mathematical pathology does not necessarily imply the removal of
the associated linear instability. In particular, regularising a model by
diffusively damping out growth at high wavenumbers can leave larger scales
unaffected and susceptible to the same dramatic instability that gave rise to
ill posedness~\citep{Baker2016,Langham2021}.  The finger-like structures
observed in granular flow fronts are a prime example of this.  Depth-averaged
equations for the dynamics of segregated bidisperse grains suffer from ill
posedness of the $O(k^{1/2})$ kind, arising from repeated real
characteristics~\citep{Woodhouse2012}.  Nevertheless, the inclusion of a
physically motivated diffusive term regularises the model and sets the preferred
fingering width~\citep{Baker2016}.  Therefore, at least in this case, ill
posedness signposts the existence of an underlying physical instability -- one
that can be correctly captured following improvements to the model.
This is more generally to be expected, since a properly formulated shallow
layer model that undergoes a Hadamard instability must be regularised with
physics that are only non-negligible over the short length scales where
the instability becomes acute. (Otherwise those terms would have to be present
in the original model formulation.)

Given the difficulties in conducting experiments and observations of debris
flows, it remains to be seen what kind of instability this analysis might be
pointing towards. Free surface instabilities that give rise to large-amplitude
`roll waves' and related phenomena are already known to occur in debris
flows~\citep{Zanuttigh2007,Schoffl2023,Chen2024}.  However, in these cases, the instability
mechanism emerges from interactions between gravitational forcing and frictional
resistance from basal stresses~\citep{Trowbridge1987}.  The instabilities that
we have considered in this paper are independent of these effects. Instead, they
arise from the coupling between the phases provided by buoyancy. Consequently,
they seem more likely to be related to interior instabilities found in
multi-layered fluid flows such as the Kelvin-Helmholtz
mechanism~\citep{Castro2001}.  In a well mixed flow of fluid and grains,
the phenomenology of such an instability would need to be quite different.
Nevertheless, perhaps it will turn out that the high frequency
resonance between the two phases is ultimately resolved similarly to the case of mixing
between fluid layers. That is, through the generation of internal vortices that
dissipate energy and act to reduce the velocity difference between the phases,
thereby driving the modelled flow away from non-hyperbolic regions. Unravelling
these issues could be an interesting challenge for future study.

\acknowledgements{
This research was supported by funding from National Environment
Research Council (NERC) grants NE/X00029X/1 and NE/X013936/1.
J.\ Langham acknowledges fruitful discussions with Andrew J.\ Hogg over the course of
this work.
X.\ Meng is grateful for the support of the China NFSC grant no.\ 12272074 and
the Liaoning Revitalization Talents Program XLYC2203149.
J.\ M.\ N.\ T.\ Gray
was supported by a Royal Society Wolfson Research Merit Award (WM150058) and an EPSRC Established Career Fellowship (EP/M022447/1).
}

~\\
Declaration of Interests.
The authors report no conflict of interest.

\appendix%

\section{Details of the figure~\ref{fig:illp example} numerical simulations}%
\label{appendix:numerics}%
The data for the illustrative simulation in figure~\ref{fig:illp example} were
obtained by numerically integrating the~\cite{Meng2022} model equations for
oversaturated debris flows [Eqs.~\eqref{eq:pitman le 1}, \eqref{eq:meng 1} for
the solids phase and Eqs.~\eqref{eq:pitman le 3}, \eqref{eq:meng 2} for the
fluid phase], using the finite volume scheme of~\cite{Kurganov2000} in
combination with the technique of \cite{Kurganov2009} to handle non-conservative
product terms.  Though  the source terms in Eqs.~\eqref{eq:meng 1}
and~\eqref{eq:meng 2} do not affect the presence of the catastrophic
instabilities in the model, they must be specified to simulate the equations.
The following closures were employed:
\begin{subequations}
\begin{gather}
    S_s = -g^x - (1-\gamma) g^z \mu_b \frac{\average{u_s}}{|\average{u_s}|}
    - \frac{C_d}{\rho_s\varphi_c}(\average{u_s}-\average{u_f}),\\
    S_f = -g^x - C_w\frac{\average{u_f}|\average{u_f}|}{H_f}
    - \frac{C_d}{\rho_f\varphi_c} \frac{H_s}{H_f}(\average{u_f}-\average{u_s}),
\end{gather}
\end{subequations}
where $\mu_b$ and $C_w$ are dimensionless coefficients and $C_d$ is dimensional
Darcy drag coefficient modelled by $C_d =
180\eta_f\varphi_c^2/[d^2(1-\varphi_c)]$, with $\eta_f$ denoting the dynamic
viscosity of the fluid and $d$ a
characteristic solids diameter.  These capture the essential
competition between downslope gravitational acceleration $g^x$, basal drag and
interphase (Darcy) drag in these systems.
The $\mu_b$ coefficient for the solids phase is dynamically set by a granular
friction law~\citep{Pouliquen2002,Jop2005}:
\begin{equation}
    \mu_b = \mu_1 + \frac{\mu_2 - \mu_1}{1 + I_0 / I},
    \quad\mathrm{where}
    \quad
    I = \frac{5|\average{u_s}|d\varphi_c^{3/2}}{2(g^z \varphi_c H_s^3)^{1/2}}
\end{equation}
is a so-called `inertial number' for the grains.

The source term parameter values used were:
$g^x = -g\sin(18.5^\circ)$, $g^z = g\cos(18.5^\circ)$, where
$g=9.8\mathrm{m}/\mathrm{s}^2$, $\rho_s = 1400\mathrm{kg}/\mathrm{m}^3$, $\rho_f
= 1000\mathrm{kg}/\mathrm{m}^3$, [implying $\gamma = 1/1.4$~($\approx 0.7$)],
$\eta_f = 10^{-3}\,\mathrm{kg}/\mathrm{m}/\mathrm{s}$,
$\mu_1 = \tan(22.5^\circ)$, $\mu_2 = \tan(30.1^\circ)$, 
$d = 8\times 10^{-3}\mathrm{m}$,
$\varphi_c = 0.5$,
$I_0 = 9/(44\sqrt{\varphi_c})$ ($\approx 0.3$)
and $C_w = 0.01$.
Additionally, diffusion of fluid momentum was neglected, i.e.\ $\nu_f = 0$
(though note that dynamic viscosity $\eta_f$ retains a nonzero value for the purposes of
the Darcy drag closure).
Simulations were conducted in a domain of length $L = 0.3\mathrm{m}$ with periodic
boundary conditions enforced for all fields at $x = 0\mathrm{m} \equiv
0.2\mathrm{m}$ and three numerical grid spacings $\Delta x = 5 \times 10^{-4}\mathrm{m}$,
$5 \times 10^{-5}\mathrm{m}$, $5 \times 10^{-6}\mathrm{m}$. 
In each case the initial condition used was a steady uniform flow in an
ill-posed regime of the model.  Such states occur when the source terms vanish,
implying flow at equilibrium, with $S_s = S_f = 0$.  Specifically, $h_s =
0.0945794565\mathrm{m}$, $\average{u_s} = 6.5195983137\mathrm{m}/\mathrm{s}$,
$h_f = 0.1176076626\mathrm{m}$, $\average{u_f} =
5.711201893\mathrm{m}/\mathrm{s}$, were set at $t = 0$. The equivalent partial depths
$H_s$, $H_f$ are obtained via the transformations in Eq.~\eqref{eq:meng
transformations}. To $3$~s.f., the corresponding dimensionless field variables
are $R_H = 0.673$, $R_u = 1.14$ and $\Fr = 7.06$.
Additionally, a small disturbance was given to this initial condition.
Specifically, each field
$q$ was initialised at $t = 0$, to the real part of
\begin{equation}
    q_0\left[1 + 
    \frac{\epsilon}{||\vect{\xi}||} \sum_{n=1}^{n=N} A_n \exp\left(\im 2\pi n x / L\right)
    \right]\!,
    \label{eq:perturbation}%
\end{equation}
where $q_0$ denotes the corresponding steady uniform flow value for the field, 
$\epsilon = 10^{-6}$, $N = L / \Delta x$ is the number of
simulation grid cells and $\vect{\xi}$ is a vector of complex-valued random
amplitudes uniformly distributed within in the unit circle, with norm
$||\vect{\xi}||=(|\xi_1|^2+\ldots+|\xi_N|^2)^{1/2}$.

\section{Table of notation}%
\label{appendix:notation}%
To ease comparison between different models and our analysis,
table~\ref{tab:symbols} lists the main symbols used in the paper, alongside the
equivalent quantities in~\cite{Pitman2005}, \cite{Pudasaini2012}
and~\cite{Meng2022} using the original authors' notation.
\begin{table}
  \begin{center}
\def~{\hphantom{0}}
  \begin{tabular}{lcccc}
      \vspace{0.1cm}
      ~ & This paper & \cite{Pitman2005} & \cite{Pudasaini2012} &
      \cite{Meng2022}\\
      Velocity & $\vect{u}_s$, $\vect{u}_f$ &
      $\vect{v}$, $\vect{u}$ & $\vect{u}_s$, $\vect{u}_f$ &
      $\vect{u}^g$, $\vect{u}^w$ \\
      Density & $\rho_s$, $\rho_f$ & $\rho^s$, $\rho^f$ &
      $\rho_s$, $\rho_f$ & $\rho^{g\star}$, $\rho^{w\star}$ \\
      Density ratio $(\rho_f/\rho_s)$ & $\gamma$ & --- & $\gamma$ & $\gamma$ \\
      Volume fraction & $\varphi_s$, $\varphi_f$ & $\varphi$,
      $1-\varphi$ & $\alpha_s$, $\alpha_f$ & $\phi^g$, $\phi^w$ \\
      Constant solids volume fraction & $\varphi_c$ & --- & --- & $\phi^c$ \\
      Effective solids stress & $\matrgr{\sigma}_s$ &
      $-\matr{T}^s$ & $\alpha_s(p\matr{I}-\matr{T}_s)$
      & $-\matrgr{\sigma}^e$ \\
      Effective fluid stress &  $\matrgr{\sigma}_f$ &
       $-\matr{T}^f$ &  $-p\matr{I}+\alpha_f\matrgr{\tau}_f$
      &  $-p^{w\star}\matr{I}+\matrgr{\tau}^w$ \\
      Pore fluid pressure & $p$ & --- & $p$ & $p^{w\star}$ \\
      Total interphase force & $\vect{f}_s$, $\vect{f}_f$ & --- & --- & --- \\
      Non-buoyant interphase force & $\vect{d}_s$, $\vect{d}_f$ &
      $\vect{f}$, $-\vect{f}$ &
      $\vect{M}_s$, $\vect{M}_f$ & --- \\
      Gravity vector & $-\vect{g}$ & $\vect{g}$
      & $\vect{f}$ & $\vect{g}$ \\
      Total flow depth & $h$ & $\hat{h}$ & $h$ & $h^w$ \\
      Partial depth & $H_s$, $H_f$ & $\average{\varphi} \hat{h}$,
      $(1-\average{\varphi})\hat{h}$ &
      $\average{\alpha_s} h$, $\average{\alpha_f} h$ & $\varphi_c h^g$,
      $h^w-\varphi_c h^g$ \\
      Solid/fluid layer depth & $h_s$, $h_f$ & --- & --- & $h^g$, $h^w$ \\
      Added mass coefficient & $C$ & --- & $C_{\mathit{VM}}$ & --- \\
      Earth pressure coefficient & $K$ & $\alpha_{xx}$ & $K_x$ & $1$ \\
      Dynamic fluid viscosity & $\eta_f$ & --- & $\eta_f$ & $\eta^w$ \\
      Non-Newtonian coefficient & $\mathcal{A}$ & --- & $\mathcal{A}$ & --- \\
  \end{tabular}
      \caption{Comparison of notation for the main two-phase models considered
      herein. 
      Where no direct analogue of a quantity exists in a given article,
      we either derive it in the authors' original notation, or 
      leave the entry blank.
      Pairs of quantities refer to solids and fluid phase components
      respectively. 
      In some cases, we retain hats and overbars that are
      eventually dropped for brevity in the original articles.
      As in the main text, the \cite{Meng2022} model is assumed to be in its
      oversaturated configuration.
      }%
  \label{tab:symbols}
  \end{center}
\end{table}
Not all the symbols can be directly translated, either because
some terms only appear in a subset of models, or due to conceptual
differences in approach.
For example, instead of the quantities that we term the `effective stresses', some
authors define stress tensors that incorporate part of the buoyancy effect
\citep[which itself is not uniquely defined in this context,
see][]{Jackson2000}.
These differences in bookkeeping, though conceptually meaningful, do not
ultimately lead to incompatible physical descriptions once the models are
carefully depth-averaged.

\section{\cite{Pudasaini2019} coefficient matrices}%
\label{appendix:3 phase}%
The analyses of \S\ref{sec:three phase} and \S\ref{sec:three phase diffusive}
investigate the eigenstructure of the frozen coefficient problem~\eqref{eq:full
eigenproblem} for the model of~\cite{Pudasaini2019}.  The underlying model
equations are lengthy and fully specified in the original paper. To obtain the
relevant matrices for our analysis, the same essential steps are followed as for
the two-phase systems.  The original equations in conservative form are
rewritten in the quasilinear form of Eq.~\eqref{eq:gen model eqs} and
non-dimensionalised with respect to the height and velocity of third (fluid)
phase, as described in the text around Eqs.~\eqref{eq:3 phase nondim}.  Then, the
coefficients are frozen around a base state given by $H_1 = R_{H_1}$,
$\average{u_1} = R_{u_1}$, $H_2 = R_{H_2}$, $\average{u_2} = R_{u_2}$, $H_3 =
\average{u_3} = 1$. Finally, the added mass coefficients that appear in the
model are assumed to be zero, implying that $\matr{A}=\matr{I}$.  The Jacobian matrix
$\matr{B}$ is constructed in \S\ref{sec:three phase}, by evaluating
Eq.~\eqref{eq:3 phase jacobian} and substituting the particular closures for
this model, which are given
in Eqs.~\eqref{eq:3 phase jac closures}.
Since it is not relevant for our analysis, there is no need to specify the
source matrix $\matr{C}$.

Newtonian and non-Newtonian stresses, analogous to those in Eq.~\eqref{eq:pudasaini
diffusion}, are included for both the fluid phases~$2$ and~$3$. This means there
are two `kinematic' viscosities, $\nu_2$ and $\nu_3$ respectively, for the
Newtonian stresses, which we render dimensionless with respect to
$H_3^{(0)}\average{u_3}^{(0)}$\!. Furthermore, a single downslope non-Newtonian
diffusive term is proposed for phase~$2$, while two such terms appear in the
momentum equation for phase~$3$~\citep{Pudasaini2019}.  This introduces three
further parameters $\mathcal{N}_{21}$, $\mathcal{N}_{31}$, $\mathcal{N}_{32}$,
which are defined similarly to the parameter $\mathcal{N}$ of
\S\ref{sec:pudasaini diffusive}, as ratios between non-Newtonian and Newtonian
diffusion coefficients.  The nonzero entries $\mathsfi{D}_{ij}$ of the diffusion
matrix $\matr{D}$ are given by
\begin{subequations}
\begin{gather}
    \mathsfi{D}_{41} =
    2\nu_2\mathcal{N}_{21}\frac{(R_{u_1}-R_{u_2})(1+R_{H_2})}{(1+R_{H_1}+R_{H_2})^2},\tag{\theequation\emph{a}}\\
    \mathsfi{D}_{43} =
    \mathsfi{D}_{45} =
        2\nu_2\mathcal{N}_{21}\frac{(R_{u_2}-R_{u_1})R_{H_1}}{(1+R_{H_1}+R_{H_2})^2},~~
    \mathsfi{D}_{44} =
    2\nu_2,\tag{\theequation\emph{b,c}}\\
    \mathsfi{D}_{61} = 
    \frac{2\nu_3}{(1+R_{H_1}+R_{H_2})^2}\left[
        \mathcal{N}_{31}(R_{u_1}-1)(1+R_{H_2}) + \mathcal{N}_{32}(1-R_{u_2})R_{H_2}
        \right]\!,\tag{\theequation\emph{d}}\\
    \mathsfi{D}_{63} = 
    \frac{2\nu_3}{(1+R_{H_1}+R_{H_2})^2}\left[
        \mathcal{N}_{31}(1-R_{u_1})R_{H_1} +
        \mathcal{N}_{32}(R_{u_2}-1)(1+R_{H_1})
        \right]\!,\tag{\theequation\emph{e}}\\
    \mathsfi{D}_{65} = 
    \frac{2\nu_3}{(1+R_{H_1}+R_{H_2})^2}\left[
        \mathcal{N}_{31}(1-R_{u_1})R_{H_1} +
        \mathcal{N}_{32}(1-R_{u_2})R_{H_2}
        \right]\!,~~
    \mathsfi{D}_{66} =
    2\nu_3.\tag{\theequation\emph{f,g}}%
\end{gather}
\end{subequations}

A convenient basis change matrix $\matr{P}$ that diagonalises $\matr{D}$ is
given by the matrix whose only nonzero entries are
\begin{subequations}
\begin{gather}
    \mathsfi{P}_{41} = -\frac{\mathcal{N}_{21}(1 + R_{H_2})(R_{u_1} -
    R_{u_2})}{(1 + R_{H_1} + R_{H_2})^2}, \quad
    \mathsfi{P}_{43} = \mathsfi{P}_{45} = \frac{\mathcal{N}_{21}(R_{u_1} -
    R_{u_2})R_{H_1}}{(1 + R_{H_1} + R_{H_2})^2}, \tag{\theequation\emph{a,b}}\\
    \mathsfi{P}_{61} =
    \frac{-\mathcal{N}_{31}(R_{u_1}-1)(1+R_{H_2})+\mathcal{N}_{32}(R_{u_2}-1)R_{H_2}}{(1
    + R_{H_1} + R_{H_2})^2},\tag{\theequation\emph{c}}\\
    \mathsfi{P}_{63} =
    \frac{\mathcal{N}_{31}(R_{u_1}-1)R_{H_1}-\mathcal{N}_{32}(R_{u_2}-1)(1+R_{H_1})}{(1
    + R_{H_1} + R_{H_2})^2},\tag{\theequation\emph{d}}\\
    \mathsfi{P}_{65} = \frac{\mathcal{N}_{31}(R_{u_1} - 1)R_{H_1} +
    \mathcal{N}_{32}(R_{u_2} - 1)R_{H_2}}{(1 +
    R_{H_1} + R_{H_2})^2} \tag{\theequation\emph{e}}%
\end{gather}
\end{subequations}
and $\mathsfi{P}_{ii} = 1$ for all $i=1,\ldots 6$.
This matrix is constructed so that the nonzero entries of $\hat{\matr{D}} =
\matr{P}^{-1}\matr{D}\matr{P}$ are $\mathsfi{D}_{55} = 2\nu_2$ and
$\mathsfi{D}_{66}=2\nu_3$.
Consequently, the reduced Jacobian $\hat{\matr{B}}_{\mathrm{red}}$ is formed by
deleting rows and columns $5$ and $6$ of $\matr{P}^{-1}\matr{B}\matr{P}$.  Its
eigenvalues are computed numerically in figure~\ref{fig:pm computations} for
various flow states, using the following illustrative model parameter values:
$\gamma_1 = \gamma_2 = 0.5$, $K = 1$, and $\mathcal{N}_{21} = \mathcal{N}_{31} =
\mathcal{N}_{32} = 5000$.
Note that since $\nu_2$ and $\nu_3$ do not appear in $\matr{P}$, 
these values do not need to be specified to reproduce
figure~\ref{fig:pm computations}.


\begin{thebibliography}{64}
\expandafter\ifx\csname natexlab\endcsname\relax\def\natexlab#1{#1}\fi
\def\au#1{#1} \def\ed#1{#1} \def\yr#1{#1}\def\at#1{#1}\def\jt#1{\textit{#1}}
  \def\bt#1{#1}\def\bvol#1{\textbf{#1}} \def\vol#1{#1} \def\pg#1{#1}
  \def\publ#1{#1}\def\arxiv#1{#1}\def\org#1{#1}\def\st#1{\textit{#1}}

\bibitem[Abgrall \& Karni(2009)]{Abgrall2009}
{\sc \au{Abgrall, R.} \& \au{Karni, S.}} \yr{2009}  \at{Two-layer shallow water
  system: a relaxation approach}.  \jt{SIAM J. Sci. Comput.}  \bvol{31}~(3),
  \pg{1603--1627}.

\bibitem[Anderson \& Jackson(1967)]{Anderson1967}
{\sc \au{Anderson, T.~B.} \& \au{Jackson, R.}} \yr{1967}  \at{Fluid mechanical
  description of fluidized beds. equations of motion}.  \jt{Ind. Eng. Chem.
  Fund.}  \bvol{6}~(4),  \pg{527--539}.

\bibitem[Baker {\em et~al.\/}(2016)Baker, Johnson \& Gray]{Baker2016}
{\sc \au{Baker, J.~L.}, \au{Johnson, C.~G.} \& \au{Gray, J. M. N.~T.}}
  \yr{2016}  \at{Segregation-induced finger formation in granular free-surface
  flows}.  \jt{J. Fluid Mech.}  \bvol{809},  \pg{168--212}.

\bibitem[Bedford \& Drumheller(1983)]{Bedford1983}
{\sc \au{Bedford, A.} \& \au{Drumheller, D.~S.}} \yr{1983}  \at{Theories of
  immiscible and structured mixtures}.  \jt{Int. J. Eng. Sci.}  \bvol{21}~(8),
  \pg{863--960}.

\bibitem[Berti {\em et~al.\/}(2000)Berti, Genevois, La{H}usen, Simoni \&
  Tecca]{Berti2000}
{\sc \au{Berti, M.}, \au{Genevois, R.}, \au{La{H}usen, R.}, \au{Simoni, A.} \&
  \au{Tecca, P.~R.}} \yr{2000}  \at{Debris flow monitoring in the {A}cquabona
  watershed on the {D}olomites ({I}talian {A}lps)}.  \jt{Phys. Chem. Earth Pt.
  B}  \bvol{25}~(9),  \pg{707--715}.

\bibitem[Bouchut {\em et~al.\/}(2016)Bouchut, Fern{\'a}ndez-Nieto, Mangeney \&
  Narbona-Reina]{Bouchut2016}
{\sc \au{Bouchut, F.}, \au{Fern{\'a}ndez-Nieto, E.~D.}, \au{Mangeney, A.} \&
  \au{Narbona-Reina, G.}} \yr{2016}  \at{A two-phase two-layer model for
  fluidized granular flows with dilatancy effects}.  \jt{J. Fluid Mech.}
  \bvol{801},  \pg{166--221}.

\bibitem[Brufau {\em et~al.\/}(2000)Brufau, Garcia-Navarro, Ghilardi, Natale \&
  Savi]{Brufau2000}
{\sc \au{Brufau, P.}, \au{Garcia-Navarro, P.}, \au{Ghilardi, P.}, \au{Natale,
  L.} \& \au{Savi, F.}} \yr{2000}  \at{1{D} mathematical modelling of debris
  flow}.  \jt{J. Hydraul. Res.}  \bvol{38}~(6),  \pg{435--446}.

\bibitem[Castro {\em et~al.\/}(2001)Castro, Mac{\'\i}as \&
  Par{\'e}s]{Castro2001}
{\sc \au{Castro, M.}, \au{Mac{\'\i}as, J.} \& \au{Par{\'e}s, C.}} \yr{2001}
  \at{A {Q}-scheme for a class of systems of coupled conservation laws with
  source term. application to a two-layer 1-{D} shallow water system}.
  \jt{ESAIM-Math. Model. Num.}  \bvol{35}~(1),  \pg{107--127}.

\bibitem[{Castro D{\'\i}az} {\em et~al.\/}(2023){Castro D{\'\i}az},
  Fern{\'a}ndez-Nieto, Garres-D{\'\i}az \& de~Luna]{CastroDiaz2023}
{\sc \au{{Castro D{\'\i}az}, M.~J.}, \au{Fern{\'a}ndez-Nieto, E.~D.},
  \au{Garres-D{\'\i}az, J.} \& \au{de~Luna, T.~M.}} \yr{2023}  \at{Discussion
  on different numerical treatments on the loss of hyperbolicity for the
  two-layer shallow water system}.  \jt{Adv. Water Res.}  \bvol{182},
  \pg{104587}.

\bibitem[Chavarr{\'\i}as {\em et~al.\/}(2019)Chavarr{\'\i}as, Schielen,
  Ottevanger \& Blom]{Chavarrias2019}
{\sc \au{Chavarr{\'\i}as, V.}, \au{Schielen, R.}, \au{Ottevanger, W.} \&
  \au{Blom, A.}} \yr{2019}  \at{Ill posedness in modelling two-dimensional
  morphodynamic problems: effects of bed slope and secondary flow}.  \jt{J.
  Fluid Mech.}  \bvol{868},  \pg{461--500}.

\bibitem[Chavarr{\'\i}as {\em et~al.\/}(2018)Chavarr{\'\i}as, Stecca \&
  Blom]{Chavarrias2018}
{\sc \au{Chavarr{\'\i}as, V.}, \au{Stecca, G.} \& \au{Blom, A.}} \yr{2018}
  \at{Ill-posedness in modeling mixed sediment river morphodynamics}.  \jt{Adv.
  Water Res.}  \bvol{114},  \pg{219--235}.

\bibitem[Chen {\em et~al.\/}(2024)Chen, Song, Chen, Feng, Li, Zhao \&
  Zhang]{Chen2024}
{\sc \au{Chen, Q.}, \au{Song, D.}, \au{Chen, X.}, \au{Feng, L.}, \au{Li, X.},
  \au{Zhao, W.} \& \au{Zhang, Y.}} \yr{2024}  \at{The erosion pattern and
  hidden momentum in debris-flow surges revealed by simple hydraulic jump
  equations}.  \jt{Water Resour. Res.}  \bvol{60}~(11),  \pg{e2023WR036090}.

\bibitem[Chiapolino \& Saurel(2018)]{Chiapolino2018}
{\sc \au{Chiapolino, A.} \& \au{Saurel, R.}} \yr{2018}  \at{Models and methods
  for two-layer shallow water flows}.  \jt{J. Comput. Phys.}  \bvol{371},
  \pg{1043--1066}.

\bibitem[Christen {\em et~al.\/}(2010)Christen, Kowalski \&
  Bartelt]{Christen2010}
{\sc \au{Christen, M.}, \au{Kowalski, J.} \& \au{Bartelt, P.}} \yr{2010}
  \at{{RAMMS}: {N}umerical simulation of dense snow avalanches in
  three-dimensional terrain}.  \jt{Cold Reg. Sci. Technol.}  \bvol{63}~(1--2),
  \pg{1--14}.

\bibitem[Cordier {\em et~al.\/}(2011)Cordier, Le \& Morales~de
  Luna]{Cordier2011}
{\sc \au{Cordier, S.}, \au{Le, M.~H.} \& \au{Morales~de Luna, T.}} \yr{2011}
  \at{Bedload transport in shallow water models: Why splitting (may) fail, how
  hyperbolicity (can) help}.  \jt{Adv. Water Res.}  \bvol{34}~(8),
  \pg{980--989}.

\bibitem[Denissen {\em et~al.\/}(2019)Denissen, Weinhart, Voortwis, Luding,
  Gray \& Thornton]{Denissen2019}
{\sc \au{Denissen, I. F.~C.}, \au{Weinhart, T.}, \au{Voortwis, A.~Te},
  \au{Luding, S.}, \au{Gray, J. M. N.~T.} \& \au{Thornton, A.~R.}} \yr{2019}
  \at{Bulbous head formation in bidisperse shallow granular flow over an
  inclined plane}.  \jt{J. Fluid Mech.}  \bvol{866},  \pg{263--297}.

\bibitem[Dowling \& Santi(2014)]{Dowling2014}
{\sc \au{Dowling, C.~A.} \& \au{Santi, P.~M.}} \yr{2014}  \at{Debris flows and
  their toll on human life: a global analysis of debris-flow fatalities from
  1950 to 2011}.  \jt{Nat. Hazards}  \bvol{71},  \pg{203--227}.

\bibitem[Drew(1983)]{Drew1983}
{\sc \au{Drew, D.~A.}} \yr{1983}  \at{Mathematical modeling of two-phase flow}.
   \jt{Ann. Rev. Fluid Mech.}  \bvol{15}~(1),  \pg{261--291}.

\bibitem[Fraccarollo \& Papa(2000)]{Fraccarollo2000}
{\sc \au{Fraccarollo, L.} \& \au{Papa, M.}} \yr{2000}  \at{Numerical simulation
  of real debris-flow events}.  \jt{Phys. Chem. Earth Pt. B}  \bvol{25}~(9),
  \pg{757--763}.

\bibitem[George \& Iverson(2014)]{George2014}
{\sc \au{George, D.~L.} \& \au{Iverson, R.~M.}} \yr{2014}  \at{A depth-averaged
  debris-flow model that includes the effects of evolving dilatancy. {II.}
  {N}umerical predictions and experimental tests}.  \jt{P. Roy. Soc. A-Math.
  Phy.}  \bvol{470}~(2170),  \pg{20130820}.

\bibitem[Gray \& Kokelaar(2010)]{Gray2010}
{\sc \au{Gray, J.M.N.T.} \& \au{Kokelaar, B.P.}} \yr{2010}  \at{Large particle
  segregation, transport and accumulation in granular free-surface flows}.
  \jt{J. Fluid Mech.}  \bvol{652},  \pg{105--137}.

\bibitem[Gray \& Edwards(2014)]{Gray2014}
{\sc \au{Gray, J. M. N.~T.} \& \au{Edwards, A.~N.}} \yr{2014}  \at{A
  depth-averaged-rheology for shallow granular free-surface flows}.  \jt{J.
  Fluid Mech.}  \bvol{755},  \pg{503--534}.

\bibitem[Hutter {\em et~al.\/}(1994)Hutter, Svendsen \& Rickenmann]{Hutter1994}
{\sc \au{Hutter, K.}, \au{Svendsen, B.} \& \au{Rickenmann, D.}} \yr{1994}
  \at{Debris flow modeling: A review}.  \jt{Continuum Mech. Therm.}  \bvol{8},
  \pg{1--35}.

\bibitem[Iverson(1997)]{Iverson1997}
{\sc \au{Iverson, R.~M.}} \yr{1997}  \at{The physics of debris flows}.
  \jt{Rev. Geophys.}  \bvol{35}~(3),  \pg{245--296}.

\bibitem[Iverson \& Denlinger(2001)]{Iverson2001}
{\sc \au{Iverson, R.~M.} \& \au{Denlinger, R.~P.}} \yr{2001}  \at{Flow of
  variably fluidized granular masses across three‐dimensional terrain: 1.
  {C}oulomb mixture theory}.  \jt{J. Geophys. Res.-Solid Earth}
  \bvol{106}~(B1),  \pg{537--552}.

\bibitem[Iverson \& George(2014)]{Iverson2014}
{\sc \au{Iverson, R.~M.} \& \au{George, D.~L.}} \yr{2014}  \at{A depth-averaged
  debris-flow model that includes the effects of evolving dilatancy. {I}.
  {P}hysical basis}.  \jt{P. Roy. Soc. A-Math. Phy.}  \bvol{470}~(2170).

\bibitem[Jackson(2000)]{Jackson2000}
{\sc \au{Jackson, R.}} \yr{2000} {\em The dynamics of fluidized particles\/}.
  \publ{Cambridge University Press}.

\bibitem[Johnson {\em et~al.\/}(2012)Johnson, Kokelaar, Iverson, Logan,
  La{H}usen \& Gray]{Johnson2012}
{\sc \au{Johnson, C.~G.}, \au{Kokelaar, B.~P.}, \au{Iverson, R.~M.}, \au{Logan,
  M.}, \au{La{H}usen, R.~G.} \& \au{Gray, J. M. N.~T.}} \yr{2012}
  \at{Grain-size segregation and levee formation in geophysical mass flows}.
  \jt{J. Geophys. Res.-Earth}  \bvol{117}~(F1), f01032.

\bibitem[Jop {\em et~al.\/}(2005)Jop, Forterre \& Pouliquen]{Jop2005}
{\sc \au{Jop, P.}, \au{Forterre, Y.} \& \au{Pouliquen, O.}} \yr{2005}
  \at{Crucial role of sidewalls in granular surface flows: consequences for the
  rheology}.  \jt{J. Fluid Mech.}  \bvol{541},  \pg{167--192}.

\bibitem[Joseph(1990)]{Joseph2013}
{\sc \au{Joseph, D.~D.}} \yr{1990} {\em Fluid dynamics of viscoelastic
  liquids\/}.  \publ{Springer}.

\bibitem[Joseph {\em et~al.\/}(1990)Joseph, Lundgren, Jackson \&
  Saville]{Joseph1990b}
{\sc \au{Joseph, D.~D.}, \au{Lundgren, T.~S.}, \au{Jackson, R.} \& \au{Saville,
  D.~A.}} \yr{1990}  \at{Ensemble averaged and mixture theory equations for
  incompressible fluid--particle suspensions}.  \jt{Int. J. Multiphas. Flow}
  \bvol{16}~(1),  \pg{35--42}.

\bibitem[Joseph \& Saut(1990)]{Joseph1990}
{\sc \au{Joseph, D.~D.} \& \au{Saut, J.~C.}} \yr{1990}  \at{Short-wave
  instabilities and ill-posed initial-value problems}.  \jt{Theor. Comp. Fluid
  Dyn.}  \bvol{1}~(4),  \pg{191--227}.

\bibitem[Kowalski \& Mc{E}lwaine(2013)]{Kowalski2013}
{\sc \au{Kowalski, J.} \& \au{Mc{E}lwaine, J.~N.}} \yr{2013}  \at{Shallow
  two-component gravity-driven flows with vertical variation}.  \jt{J. Fluid
  Mech.}  \bvol{714},  \pg{434--462}.

\bibitem[Krvavica {\em et~al.\/}(2018)Krvavica, Tuhtan \&
  Jeleni{\'c}]{Krvavica2018}
{\sc \au{Krvavica, N.}, \au{Tuhtan, M.} \& \au{Jeleni{\'c}, G.}} \yr{2018}
  \at{Analytical implementation of roe solver for two-layer shallow water
  equations with accurate treatment for loss of hyperbolicity}.  \jt{Adv. Water
  Res.}  \bvol{122},  \pg{187--205}.

\bibitem[Kurganov \& Petrova(2009)]{Kurganov2009}
{\sc \au{Kurganov, A.} \& \au{Petrova, G.}} \yr{2009}  \at{Central-upwind
  schemes for two-layer shallow water equations}.  \jt{SIAM J. Sci. Comput.}
  \bvol{31}~(3),  \pg{1742--1773}.

\bibitem[Kurganov \& Tadmor(2000)]{Kurganov2000}
{\sc \au{Kurganov, A.} \& \au{Tadmor, E.}} \yr{2000}  \at{New high-resolution
  central schemes for nonlinear conservation laws and convection--diffusion
  equations}.  \jt{J. Comput. Phys.}  \bvol{160}~(1),  \pg{241--282}.

\bibitem[Langham {\em et~al.\/}(2021)Langham, Woodhouse, Hogg \&
  Phillips]{Langham2021}
{\sc \au{Langham, J.}, \au{Woodhouse, M.~J.}, \au{Hogg, A.~J.} \& \au{Phillips,
  J.~C.}} \yr{2021}  \at{Linear stability of shallow morphodynamic flows}.
  \jt{J. Fluid Mech.}  \bvol{916},  \pg{A31}.

\bibitem[Li {\em et~al.\/}(2018)Li, Cao, Hu, Pender \& Liu]{Li2018}
{\sc \au{Li, J.}, \au{Cao, Z.}, \au{Hu, K.}, \au{Pender, G.} \& \au{Liu, Q.}}
  \yr{2018}  \at{A depth-averaged two-phase model for debris flows over fixed
  beds}.  \jt{Int. J. Sediment Res.}  \bvol{33}~(4),  \pg{462--477}.

\bibitem[Macedonio \& Pareschi(1992)]{Macedonio1992}
{\sc \au{Macedonio, G.} \& \au{Pareschi, M.~T.}} \yr{1992}  \at{Numerical
  simulation of some lahars from {M}ount {S}t. {H}elens}.  \jt{J. Volcanol.
  Geoth. Res.}  \bvol{54}~(1-2),  \pg{65--80}.

\bibitem[Maxey \& Riley(1983)]{Maxey1983}
{\sc \au{Maxey, M.~R.} \& \au{Riley, J.~J.}} \yr{1983}  \at{Equation of motion
  for a small rigid sphere in a nonuniform flow}.  \jt{Phys. Fluids}
  \bvol{26}~(4),  \pg{883--889}.

\bibitem[McCoy {\em et~al.\/}(2010)McCoy, Kean, Coe, Staley, Wasklewicz \&
  Tucker]{Mccoy2010}
{\sc \au{McCoy, S.~W.}, \au{Kean, J.~W.}, \au{Coe, J.~A.}, \au{Staley, D.~M.},
  \au{Wasklewicz, T.~A.} \& \au{Tucker, G.~E.}} \yr{2010}  \at{Evolution of a
  natural debris flow: In situ measurements of flow dynamics, video imagery,
  and terrestrial laser scanning}.  \jt{Geology}  \bvol{38}~(8),
  \pg{735--738}.

\bibitem[Meng {\em et~al.\/}(2022)Meng, Johnson \& Gray]{Meng2022}
{\sc \au{Meng, X.}, \au{Johnson, C.~G.} \& \au{Gray, J. M. N.~T.}} \yr{2022}
  \at{Formation of dry granular fronts and watery tails in debris flows}.
  \jt{J. Fluid Mech.}  \bvol{943},  \pg{A19}.

\bibitem[Meng {\em et~al.\/}(2024)Meng, Taylor-Noonan, Johnson, Take, Bowman \&
  Gray]{Meng2024}
{\sc \au{Meng, X.}, \au{Taylor-Noonan, A.~M.}, \au{Johnson, C.~G.}, \au{Take,
  W.~A.}, \au{Bowman, E.~T.} \& \au{Gray, J. M. N.~T.}} \yr{2024}
  \at{Granular-fluid avalanches: the role of vertical structure and velocity
  shear}.  \jt{J. Fluid Mech.}  \bvol{980},  \pg{A11}.

\bibitem[Meyrat {\em et~al.\/}(2022)Meyrat, McArdell, Ivanova, M{\"u}ller \&
  Bartelt]{Meyrat2022}
{\sc \au{Meyrat, G.}, \au{McArdell, B.}, \au{Ivanova, K.}, \au{M{\"u}ller, C.}
  \& \au{Bartelt, P.}} \yr{2022}  \at{A dilatant, two-layer debris flow model
  validated by flow density measurements at the {S}wiss {I}llgraben test site}.
   \jt{Landslides}  \bvol{19}~(2),  \pg{265--276}.

\bibitem[Morland(1992)]{Morland1992}
{\sc \au{Morland, L.~W.}} \yr{1992}  \at{Flow of viscous fluids through a
  porous deformable matrix}.  \jt{Surv. Geophys.}  \bvol{13}~(3),
  \pg{209--268}.

\bibitem[Ovsyannikov(1979)]{Ovysannikov1979}
{\sc \au{Ovsyannikov, L.~V.}} \yr{1979}  \at{Two-layer ``shallow water''
  model}.  \jt{J. Appl. Mech. Tech. Phys.}  \bvol{20}~(2),  \pg{127--135}.

\bibitem[Pailha \& Pouliquen(2009)]{Pailha2009}
{\sc \au{Pailha, M.} \& \au{Pouliquen, O.}} \yr{2009}  \at{A two-phase flow
  description of the initiation of underwater granular avalanches}.  \jt{J.
  Fluid Mech.}  \bvol{633},  \pg{115--135}.

\bibitem[Pelanti {\em et~al.\/}(2008)Pelanti, Bouchut \& Mangeney]{Pelanti2008}
{\sc \au{Pelanti, M.}, \au{Bouchut, F.} \& \au{Mangeney, A.}} \yr{2008}  \at{A
  {R}oe-type scheme for two-phase shallow granular flows over variable
  topography}.  \jt{ESAIM-Math. Model. Num.}  \bvol{42}~(5),  \pg{851--885}.

\bibitem[Pierson(1995)]{Pierson1995}
{\sc \au{Pierson, T.~C.}} \yr{1995}  \at{Flow characteristics of large
  eruption-triggered debris flows at snow-clad volcanoes: constraints for
  debris-flow models}.  \jt{J. Volcanol. Geotherm. Res.}  \bvol{66}~(1),
  \pg{283--294}, models of Magnetic Processes and Volcanic Eruptions.

\bibitem[Pitman \& Le(2005)]{Pitman2005}
{\sc \au{Pitman, E.~B.} \& \au{Le, L.}} \yr{2005}  \at{A two-fluid model for
  avalanche and debris flows}.  \jt{Philos. T. R. Soc. A}  \bvol{363}~(1832),
  \pg{1573--1601}.

\bibitem[Pouliquen \& Forterre(2002)]{Pouliquen2002}
{\sc \au{Pouliquen, O.} \& \au{Forterre, Y.}} \yr{2002}  \at{Friction law for
  dense granular flows: application to the motion of a mass down a rough
  inclined plane}.  \jt{J. Fluid Mech.}  \bvol{453},  \pg{133--151}.

\bibitem[Pudasaini(2012)]{Pudasaini2012}
{\sc \au{Pudasaini, S.~P.}} \yr{2012}  \at{A general two-phase debris flow
  model}.  \jt{J. Geophys. Res.-Earth}  \bvol{117},  \pg{F03010}.

\bibitem[Pudasaini \& Mergili(2019)]{Pudasaini2019}
{\sc \au{Pudasaini, S.~P.} \& \au{Mergili, M.}} \yr{2019}  \at{A multi-phase
  mass flow model}.  \jt{J. Geophys. Res.-Earth}  \bvol{124}~(12),
  \pg{2920--2942}.

\bibitem[Sarno {\em et~al.\/}(2017)Sarno, Carravetta, Martino, Papa \&
  Tai]{Sarno2017}
{\sc \au{Sarno, L.}, \au{Carravetta, A.}, \au{Martino, R.}, \au{Papa, M.~N.} \&
  \au{Tai, Y.-C.}} \yr{2017}  \at{Some considerations on numerical schemes for
  treating hyperbolicity issues in two-layer models}.  \jt{Adv. Water Res.}
  \bvol{100},  \pg{183--198}.

\bibitem[Savage \& Hutter(1989)]{Savage1989}
{\sc \au{Savage, S.~B.} \& \au{Hutter, K.}} \yr{1989}  \at{The motion of a
  finite mass of granular material down a rough incline}.  \jt{J. Fluid Mech.}
  \bvol{199},  \pg{177--215}.

\bibitem[Sch{\"o}ffl {\em et~al.\/}(2023)Sch{\"o}ffl, Nagl, Koschuch,
  Schreiber, H{\"u}bl \& Kaitna]{Schoffl2023}
{\sc \au{Sch{\"o}ffl, T.}, \au{Nagl, G.}, \au{Koschuch, R.}, \au{Schreiber,
  H.}, \au{H{\"u}bl, J.} \& \au{Kaitna, R.}} \yr{2023}  \at{A perspective of
  surge dynamics in natural debris flows through pulse-doppler radar
  observations}.  \jt{J. Geophys. Res.-Earth}  \bvol{128}~(9),
  \pg{e2023JF007171}.

\bibitem[Shieh {\em et~al.\/}(1996)Shieh, Jan \& Tsai]{Shieh1996}
{\sc \au{Shieh, C.-L.}, \au{Jan, C.-D.} \& \au{Tsai, Y.-F.}} \yr{1996}  \at{A
  numerical simulation of debris flow and its application}.  \jt{Nat. Hazards}
  \bvol{13},  \pg{39--54}.

\bibitem[Stecca {\em et~al.\/}(2014)Stecca, Siviglia \& Blom]{Stecca2014}
{\sc \au{Stecca, G.}, \au{Siviglia, A.} \& \au{Blom, A.}} \yr{2014}
  \at{Mathematical analysis of the {S}aint-{V}enant-{H}irano model for
  mixed-sediment morphodynamics}.  \jt{Water Resour. Res.}  \bvol{50}~(10),
  \pg{7563--7589}.

\bibitem[Takahashi {\em et~al.\/}(1992)Takahashi, Nakagawa, Harada \&
  Yamashiki]{Takahashi1992}
{\sc \au{Takahashi, T.}, \au{Nakagawa, H.}, \au{Harada, T.} \& \au{Yamashiki,
  Y.}} \yr{1992}  \at{Routing debris flows with particle segregation}.  \jt{J.
  Hydraul. Eng.}  \bvol{118}~(11),  \pg{1490--1507}.

\bibitem[Trowbridge(1987)]{Trowbridge1987}
{\sc \au{Trowbridge, J.~H.}} \yr{1987}  \at{Instability of concentrated free
  surface flows}.  \jt{J. Geophys. Res.--Oceans}  \bvol{92}~(C9),
  \pg{9523--9530}.

\bibitem[Trujillo-Vela {\em et~al.\/}(2022)Trujillo-Vela, Ramos-Ca{\~n}{\'o}n,
  Escobar-Vargas \& Galindo-Torres]{Trujillo2022}
{\sc \au{Trujillo-Vela, M.~G.}, \au{Ramos-Ca{\~n}{\'o}n, A.~M.},
  \au{Escobar-Vargas, J.~A.} \& \au{Galindo-Torres, S.~A.}} \yr{2022}  \at{An
  overview of debris-flow mathematical modelling}.  \jt{Earth-Sci. Rev.}
  \bvol{232},  \pg{104135}.

\bibitem[Vreugdenhil(1979)]{Vreugdenhil1979}
{\sc \au{Vreugdenhil, C.~B.}} \yr{1979}  \at{Two-layer shallow-water flow in
  two dimensions, a numerical study}.  \jt{J. Comput. Phys.}  \bvol{33}~(2),
  \pg{169--184}.

\bibitem[Woodhouse {\em et~al.\/}(2012)Woodhouse, Thornton, Johnson, Kokelaar
  \& Gray]{Woodhouse2012}
{\sc \au{Woodhouse, M.~J.}, \au{Thornton, A.~R.}, \au{Johnson, C.~G.},
  \au{Kokelaar, B.~P.} \& \au{Gray, J. M. N.~T.}} \yr{2012}
  \at{Segregation-induced fingering instabilities in granular free-surface
  flows}.  \jt{J. Fluid Mech.}  \bvol{709},  \pg{543--580}.

\bibitem[Zanuttigh \& Lamberti(2007)]{Zanuttigh2007}
{\sc \au{Zanuttigh, B.} \& \au{Lamberti, A.}} \yr{2007}  \at{Instability and
  surge development in debris flows}.  \jt{Rev. Geophys.}  \bvol{45}~(3),
  rG3006.

\end{thebibliography}
\end{document}